\documentclass[
reprint,
superscriptaddress,
amsmath,amssymb,
aps,
pra,
twocolumn
]{revtex4-2}

\usepackage{graphicx}
\usepackage{dcolumn}
\usepackage{bm}
\usepackage[breaklinks]{hyperref}
\hypersetup{colorlinks=true, linkcolor=blue, citecolor=blue, filecolor=blue, urlcolor=blue}
\graphicspath{{figures}}
\usepackage{bm}

\usepackage{graphicx,subfigure}
\usepackage{epsfig}
\usepackage{bm}
\usepackage{dcolumn}
\usepackage{color}
\usepackage{physics}
\usepackage{float}
\makeatletter
\newcommand*{\rom}[1]{\expandafter\@slowromancap\romannumeral #1@}
\makeatother
\usepackage{lipsum}
\usepackage{subfigure}
\usepackage{ulem}
\usepackage{dsfont}
\usepackage{enumitem}
\usepackage{tikz}
\usetikzlibrary{quantikz}

\newcommand{\rz}[1]{\textcolor[rgb]{0,0.0,0}{#1}}
\newcommand{\rzz}[1]{\textcolor[rgb]{0,0.0,0}{#1}}

\begin{document}
\title{Circuit structure-preserving error mitigation for High-Fidelity Quantum Simulations}

\author{Ruizhe Shen}
\email{e0554228@u.nus.edu}
\affiliation{Department of Physics, National University of Singapore, Singapore 117542}

\author{Tianqi Chen}
\affiliation{School of Physical and Mathematical Sciences, Nanyang Technological University, Singapore 639798}

\author{Ching Hua Lee}
\email{phylch@nus.edu.sg}
\affiliation{Department of Physics, National University of Singapore, Singapore 117542}

\date{\today}

\begin{abstract}
Developing methods to accurately characterize and mitigate the impact of noise is crucial for enhancing the fidelity of quantum simulations on Noisy Intermediate-Scale Quantum (NISQ) devices. In this work, we present a circuit structure-preserving error mitigation framework for parameterized quantum circuits. A key advantage of our approach lies in its ability to retain the original circuit architecture while effectively characterizing and mitigating gate errors, enabling robust and high-fidelity simulations. This makes it particularly well suited for small-scale circuits that require repeated execution at large sampling rates. To demonstrate the effectiveness of our method, we perform variational quantum simulations of a non-Hermitian ferromagnetic transverse-field Ising chain on IBM Quantum processors. The mitigated result shows excellent agreement with exact theoretical predictions across a range of noise levels. Our strategy offers a practical solution for addressing gate-induced errors and significantly broadens the scope of feasible quantum simulations on current quantum hardware.
\end{abstract}  

\pacs{}  
\maketitle

\section{Introduction}
Recent advances in quantum computing have sparked increasing interest in utilizing noisy intermediate-scale quantum (NISQ) devices for efficient quantum simulations \cite{steane1998quantum,knill2010quantum,preskill2018quantum,national2019quantum}. Such simulations are crucial in deepening our understanding of fundamental quantum physics and exploring intriguing quantum phenomena.
NISQ platforms offer highly tunable qubit couplings and reprogrammable circuit architectures, providing exceptional flexibility for simulating a wide variety of quantum systems. This versatility makes them powerful tools with broad applications across quantum chemistry and condensed matter physics \cite{randall2021many,chen2023robust,zhou2020quantum,harrigan2021quantum,montanaro2024quantum,koh2024realization,chen2024direct,koh2022simulation,koh2022stabilizing,ippoliti2021many,xu2021realizing,brennen2008measurement,poulin2009preparing,wecker2015solving,nam2020ground,stanisic2022observing,shen2025robust,sciorilli2025towards,kirmani2022probing,agresti2024demonstration,wang2024pulse,koh2025interacting}.

Despite the promising potential of NISQ devices, achieving robustly accurate quantum simulations remains a significant challenge \cite{gingrich2004non,williams2004probabilistic,hu2020quantum,head2021capturing}.  Current quantum computers are constrained by several factors, including limited circuit depth, qubit decoherence, and gate errors  \cite{preskill2018quantum,RevModPhys.95.045005}. To mitigate these issues, various error-mitigation techniques have been developed, notably zero-noise extrapolation (ZNE) \cite{giurgica2020digital,pascuzzi2022computationally,he2020zero} and probabilistic error cancellation (PEC) \cite{mari2021extending,van2023probabilistic,gupta2024probabilistic,camilo2025compilation,}. ZNE enhances simulation accuracy by performing measurements at multiple noise levels and extrapolating results to a theoretical zero-noise limit. However, the effectiveness of ZNE is fundamentally constrained by finite qubit coherence times, limiting the range of noise variation.  PEC, on the other hand, aims to counteract noise effects by probabilistically inverting error channels. Although PEC can also produce high-fidelity outcomes, its success heavily relies on accurate noise modeling. This reliance poses significant practical challenges since noise characteristics in real quantum hardware are typically unstable, and fluctuating due to environmental variations and hardware instabilities. \rz{Although ZNE have demonstrated notable improvements in noisy quantum simulations, they rely heavily on predefined noise models and, crucially, might require significant modifications to circuit structures to perform noise characterization. Due to such extensive circuit modification, these error mitigation methods might not provide comprehensive and precise characterization of noise impacts across systems. This limitation underscores an urgent need for more adaptive, structure-preserving, and robust error-mitigation strategies.} \rzz{Moreover, in quantum simulations, there exist important classes of tasks that do not rely on large or deeply structured circuits but instead require the repeated execution of small-scale circuits a very large number of times. In such cases, the challenge lies not in scaling up circuit depth but in maintaining robustness and consistency across many repetitions on noisy hardware. Developing methods that enhance the reliability of these high-repetition, small-circuit jobs is therefore crucial for advancing near-term quantum applications.}

Thus, we introduce an efficient framework of circuit structure-preserving error mitigation, which can characterize noise impact without extensive circuit modifications. Our method employs a copy of the original quantum circuit to extract noise information while preserving its circuit structure, enabling precise and consistent noise characterization.  By integrating calibration circuits constructed by this characterization, we 
 demonstrate that our method can characterize and suppress noise effects across the entire quantum circuit \cite{mooney2021whole, nation2021scalable}. We then validate our approach by simulating long-time dynamics of a non-Hermitian transverse-field Ising chain using parameterized circuits optimized through variational methods \cite{zhu2019training,google2020hartree,cerezo2021variational,ostaszewski2021structure,chen2022high,koh2023observation}. Our noisy simulations on IBM Quantum devices show excellent agreement with exact classical results, demonstrating the robustness and efficiency of our enhanced mitigation strategy.  Overall, our error-mitigation framework offers a promising pathway for performing reliable variational quantum simulations, with broad applicability to quantum phenomena such as nonequilibrium dynamics, nonunitary physics, and ground-state characterization \cite{shen2022non,qin2022universal,liu2021non,yang2024non,qin2024kinked,zhang2022universal,kawabata2022many,jiang2022dimensional,yoshida2023fate,poddubny2023interaction,shen2025non,fu2023anatomy,li2025phase,shen2023proposal,PhysRevX.5.041003,PhysRevLett.121.237401,qin2025dynamical,lapierre2025driven}. In particular, we demonstrate that our method is especially well suited for tasks involving small-scale circuits that require a large number of circuit executions. This makes it highly relevant for applications where repeated sampling is essential, such as quantum neural network (QNN) implementations on NISQ devices \cite{schuld2014quest,jeswal2019recent,altaisky2001quantum,ricks2003training}. 

This work is structured as follows. In Section~\ref{sec2}, we introduce our error-mitigation technique and discuss a variational framework to implement our method. In Section~\ref{sec3}, we first introduce our simulation model and evaluate the performance of our error mitigation method using classical noisy simulation benchmarks. Building on these results, we then present results from quantum simulations conducted on IBM Q hardware in Section~\ref{sec4}, demonstrating that our approach significantly improves simulation fidelity. Finally, in Section~\ref{sec5}, we summarize our key contributions and discuss the broader applicability of our method.

\section{Quantum circuits and enhanced error-mitigation strategy}\label{sec2}

\subsection{Circuit structure--preserving error mitigation }

\rz{In this section, we first introduce our structure-preserving error mitigation strategy. Traditional error mitigation approaches generally rely on inserting additional components to stabilize noise or modifying quantum circuits to probe and infer noise characteristics from their responses \cite{giurgica2020digital,pascuzzi2022computationally,he2020zero,mari2021extending,van2023probabilistic}. However, such modifications can impact inherent circuit structure, making it challenging to precisely and reliably capture noise profile. Our method addresses this limitation by enabling noise characterization without altering the original circuit structure. By preserving the circuit architecture, our approach provides a faithful modeling of inherent noise.} \rzz{Moreover, we provide a general description of the simulation cost on a quantum platform, which can be parameterized by the number of qubits $L$ and the number of circuit executions $N$ (with each execution involving thousands of circuit runs): $\text{Cost}=L\times N$. In some cases, small circuits must be executed at very high volume, leading to costs comparable to those of larger circuits with fewer executions. This highlights the importance of developing an error-mitigation framework that can be efficiently reused across repeated circuit executions.}

Then, we introduce a general framework of our method. In noisy and noiseless quantum circuits, measurements in the computational basis lead to probability distributions, which can be related by a linear transformation. To describe this relationship more generally, we here propose a relationship between outputs of noisy and noiseless executions of a given quantum circuit $V$, represented as a linear transformation:
\begin{equation}\label{m}
	M(V_{\rm noiseless}\ket{\psi^{\rm in}_{i}})=V_{\rm noisy}\ket{\psi^{\rm in}_{i}}.
\end{equation}
where $\ket{\psi^{\rm in}_{i}}$ denotes an input product state from the computational basis \cite{nation2021scalable}.  The term $V_{\rm noisy/noiseless}$ refers to the noisy/noiseless implementation of the circuit $V$.  Here, the calibration matrix $M$ captures the impact of noise by mapping the ideal circuit outputs to their corresponding noisy counterparts \cite{rodriguez2022real,robertson2022mitigating,Qiskit}. In general, constructing the calibration matrix requires evaluating circuits on all basis input states, allowing for a comprehensive characterization of errors across the full computational space.

Next, we illustrate how to extract noise information without modifying the circuit structure of $V$.
To achieve this, we require the circuit $V$ to be parameterized, such that its architecture remains fixed while gate parameters can be varied and optimized for different simulations. Leveraging this structure, we make the key assumption that preserving the circuit architecture while varying only parameters can result in a consistent noise profile across executions. Mathematically, this is expressed as:
\begin{equation}
 {V}_{\text {noisy}}(\boldsymbol{\theta}) \approx \mathcal{N}\mathcal{V}(\boldsymbol{\theta}),
\end{equation}
where $\mathcal{V}(\boldsymbol{\theta})$  denotes the ideal unitary component parameterized by $\boldsymbol{\theta}$, and $\mathcal{N}$ is a parameter-independent noise channel for gate and measurement errors.  This approximation is well justified by the hardware-level implementation of gates. Most gates are implemented using analog control pulses. Varying parameters, such as rotation angles, generally alters only the pulse amplitude or duration, without affecting gate type or hardware execution path. As long as these parameter changes remain within standard operating ranges, the associated noise profile can be reasonably assumed to remain consistent across different circuit executions.

Based on this assumption, we introduce a structure-preserving calibration circuit $V^{\rm mit}$ to characterize the noise in the original circuit $V$. This leads us to propose the following procedure to model the noise effect in the original process described by Eq.~\ref{m}:
\begin{equation}\label{mit}
	M^{\rm mit}(V^{\rm mit}_{\rm noiseless}\ket{\psi^{\rm in}_{i}})=(V^{\rm mit}_{\rm noisy}\ket{\psi^{\rm in}_{i}}),
\end{equation}
where the identity circuit $V^{\rm mit}_{\rm noiseless}$ shares an identical structure with the target simulation circuit $V$. $V^{\rm mit}_{\rm noisy}$ represents the implementation of this calibration circuit on a noisy quantum device or simulator. By characterizing the noise in this structure-preserving identity circuit, we can build an approximate mitigation model for the original circuit $V^{\rm mit}$ without altering its structure.

\begin{figure}
	\centering
	\includegraphics[width=0.99\linewidth]{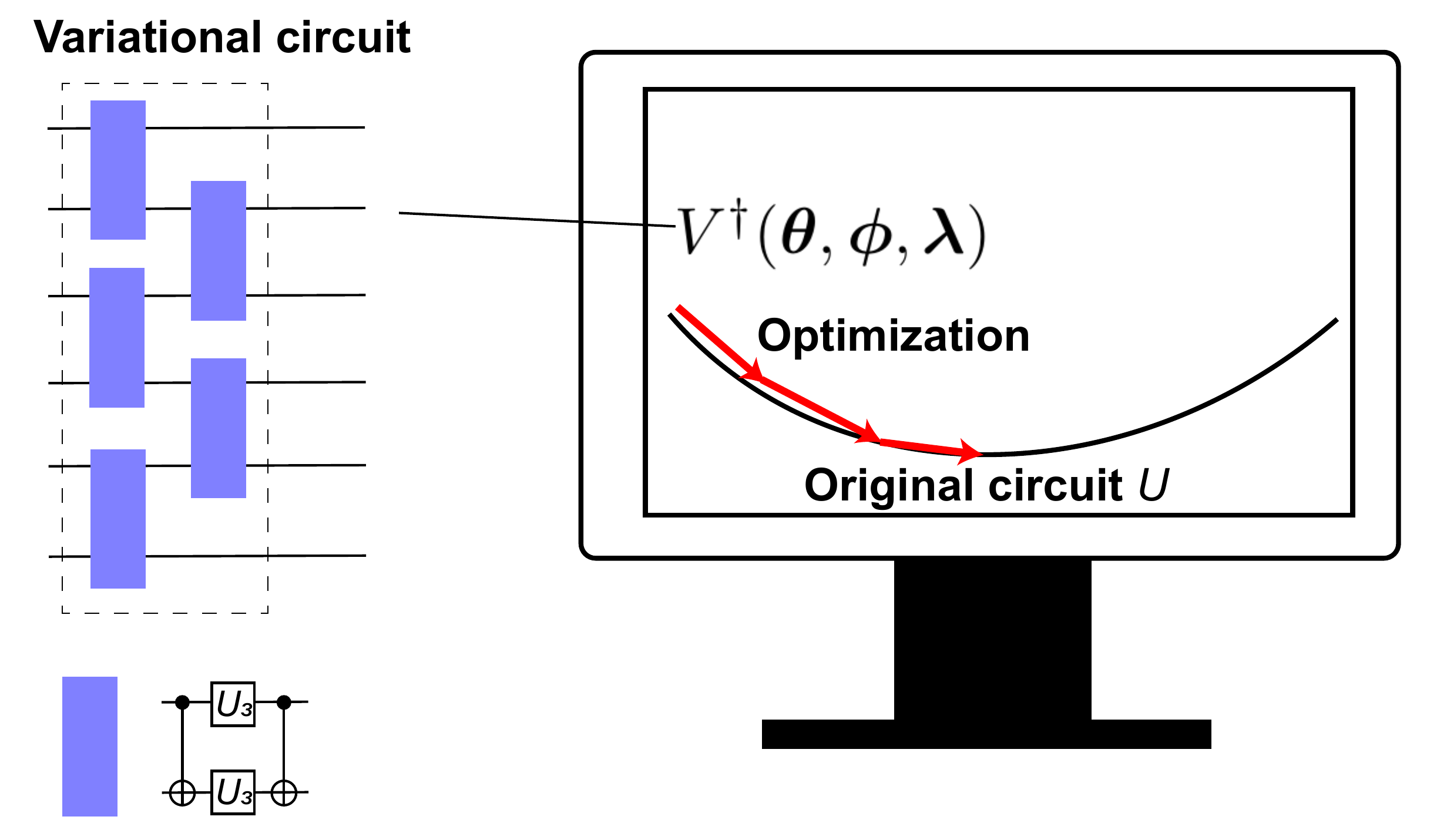}
	\caption{Schematic illustration of the variational framework for quantum simulations. A parameterized ansatz circuit $V$ is designed to approximate the unitary operator denoted by $U$ (Eq.~\eqref{u} in Appendix). A single layer, comprising parameterized $U_{3}$ gates and CX gates, is highlighted in the blue box, and the complete variational circuit consists of $n$ layers. Variational optimization is performed on a classical computer and is trained by minimizing the cost function defined in Eq.~\eqref{c}. }
	\label{fig:mps}
\end{figure}

\begin{figure*}
	\centering
	\includegraphics[width=0.8\linewidth]{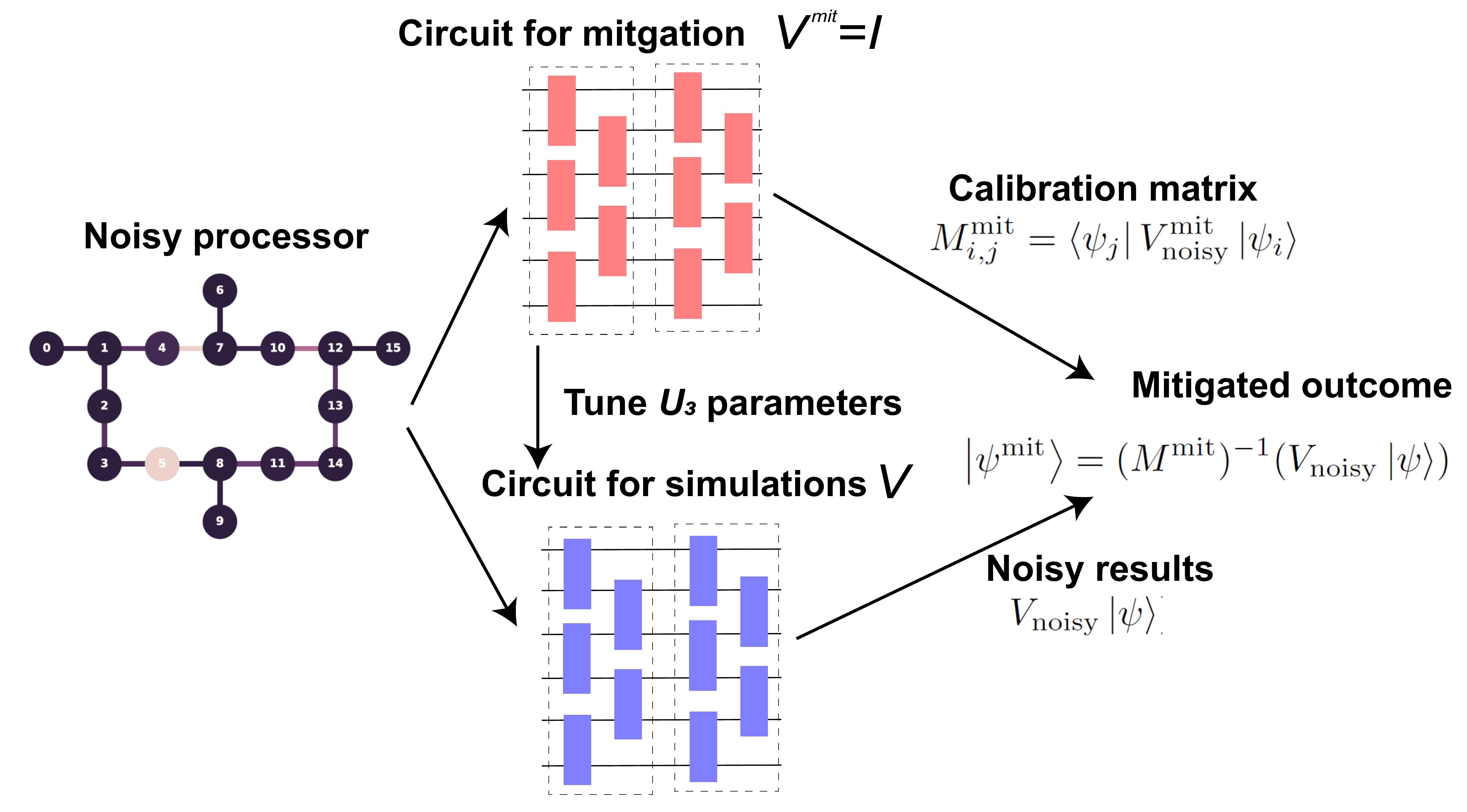}
	\caption{Schematic illustration of our structure-preserving error mitigation method applied to variational circuits. Our method consists of two key steps. First, a parameterized ansatz circuit in red for error mitigation $V^{\rm mit}$ is trained to approximate an identity operation $I$ through the optimization procedure outlined in FIG.\ref{fig:mps}. By executing all such mitigation circuits across the computational basis, the corresponding calibration matrix defined in Eq.\eqref{m}  is constructed to capture the noise characteristics of quantum devices. In the second step, the structure of this ansatz circuit is preserved while the parameters of  $U_{3}$ are re-optimized to simulate the dynamics of target models. The final error-mitigated results are obtained by applying the procedure described in Eq.~\ref{mitres}.}
	\label{fig:error}
\end{figure*}
To simplify implementation, we define the calibration circuit as the identity operation, $V^{\rm mit}_{\rm noiseless}=I$, and this identity circuit $V^{\rm mit}$ can be easily constructed by circuit optimization. With this setup, the mitigation procedure can be simplified as follows:
\begin{equation}\label{mit2}
	M^{\rm mit}\ket{\psi^{\rm in}_{i}}=V^{\rm mit}_{\rm noisy}\ket{\psi^{\rm in}_{i}},
\end{equation}
where the noiseless identity operation $V^{\rm mit}_{\rm noiseless}$ is omitted, as it leaves the input state unchanged.   Elements of the calibration matrix are then obtained as:
\begin{equation}\label{mit3}
	M^{\rm mit}_{i,j}=\bra{\psi_{j}}V^{\rm mit}_{\rm noisy}\ket{\psi_{i}},
\end{equation}
where $\ket{\psi_{i}}$ is for a product state in computational basis. Thus, this matrix, which almost embeds the full noise profile of the circuit $V$, captures the impact of all gate errors by mapping each ideal input state to its corresponding noisy output.

By collecting all noisy outcomes $(V^{\rm mit}_{\rm noisy}\ket{\psi^{\rm in}_{i}})$ for all initial states $\ket{\psi^{\rm in}_{i}}$, spanning the entire computational basis,  we systematically construct the complete calibration matrix $M^{\rm mit}$ \cite{nation2021scalable}. Assuming that the circuit $V^{\rm mit}_{\rm noisy}$ and its noisy execution $V_{\rm noisy}$ experience a similar noise channel, we can then correct noisy outcomes using the inverse of the calibration matrix:
\begin{equation}\label{mitres}
	\ket{\psi^{\rm mit}}=(M^{{\rm mit}})^{-1}(V_{\rm noisy}\ket{\psi})
\end{equation}
where $\ket{\psi}$ can be any initial state. In the following section, we present a concrete example of parameterized circuits to demonstrate the practical implementation and effectiveness of our structure-preserving error mitigation method.

\subsection{Implementation in variational circuits}
A key feature of our method is that it enables noise characterization without modifying circuit structures.
This can be achieved by employing parameterized quantum circuits, which support flexible optimization while preserving a fixed gate architecture.  By reusing the same circuit layout and varying only internal parameters, our method enables approximately consistent noise modeling across different tasks. This structure-preserving property makes our method especially well-suited for variational frameworks, where trainable circuits are optimized for a wide range of quantum simulations. Thus, in the following, we adopt this variational framework to illustrate our structure-preserving error mitigation technique.

In our work, we employ variational circuits to simulate a unitary operator $U$, specifically defined by Eq.\ref{u} in Appendix.~\ref{apsec1}. To approximate this target operator, we construct a parameterized quantum circuit and optimize it using variational techniques \cite{Qiskit,yang2017optimizing,van2021measurement,PhysRevA.111.042616,ji2025algorithm,wang2024pulse,cerezo2021variational,bittel2021training,wang2021noise,ferreira2025variational,du2022efficient}. This variational approach is highly flexible and applicable to a broad spectrum of quantum systems, extending well beyond condensed matter physics. Given the extensive applicability of the variational framework, effective mitigation of gate errors in variational circuits is crucial to achieving high-fidelity quantum simulations.

Our variational framework is presented in FIG.\ref{fig:mps}. Here,  we highlight a single layer of our ansatz circuit, composed of $U_{3}$ and  CX gates, and the full circuit consists of $n$ such layers. The optimization process is also outlined in FIG.\ref{fig:mps}. Initially, we prepare the target operator $U$ and construct a parameterized ansatz circuit $V$ using $U_{3}$ and  CX  gates. As shown in FIG.\ref{fig:mps}, each blue block acting on $i$ and $i+1$ qubits represents the operation $\text{CX}^{i,i+1}U^{i}_{3}U^{i+1}_{3}\text{CX}^{i,i+1}$. Each $U_{3}$ gate is parameterized by three angles: $\theta$, $\phi$, and $\lambda$, and is defined as
\begin{equation}
	U_{3}(\theta, \phi, \lambda)=\left[\begin{array}{cc}
		\cos \left(\frac{\theta}{2}\right) & -e^{i \lambda} \sin \left(\frac{\theta}{2}\right) \\
		e^{i \phi} \sin \left(\frac{\theta}{2}\right) & e^{i(\phi+\lambda)} \cos \left(\frac{\theta}{2}\right)
	\end{array}\right].
\end{equation}

The training process involves optimizing parameters of $U_{3}$ gates  by minimizing the following cost function:
\begin{equation}\label{c}
	C({\bm \theta, \bm\phi, \bm\lambda})=1-|\bra{\psi_{0}}V^{\dagger}U\ket{\psi_{0}}|,
\end{equation}
where $\ket{\psi_{0}}$ represents an initial state in our simulation. For our simulations presented in the main text, the circuit $U$ consists of 5 qubit in total.  In our work, such optimization can be achieved through the L-BFGS-B algorithm \cite{zhu1997algorithm}, which achieves reasonable convergence with as few as two circuit layers. A more detailed discussion of this optimization strategy is presented in Appendix~\ref{apvqa}.

The workflow of our error-mitigation procedure for the above variational simulation is illustrated in FIG.~\ref{fig:error}. To ensure consistency in noise characterization, we integrate the mitigation and simulation circuits $V^{\rm mit}$ and $V$ with identical structures, which are both built by $n$ layers and execute them simultaneously on quantum hardware. This approach minimizes variations in noise between different circuit runs. The final mitigated result is then obtained by applying the inverse calibration matrix, as described in Eq.\ref{mitres}.

\section{Demonstration of enhanced error mitigation}\label{sec3}
\subsection{Illustrative model
}\label{sec1}

\rz{To demonstrate the effectiveness of our method, we simulate the dynamical evolution of a quantum system using variational circuits. Recent studies have shown that variational approaches are particularly well-suited for capturing nonunitary dynamics in non-Hermitian quantum models \cite{shen2025observation,sun2021quantum,kamakari2022digital,mcardle2019variational,motta2020determining}.}  Motivated by these developments, we focus on a prototypical example: a non-Hermitian transverse-field Ising (TFI) chain  \cite{shibata2019dissipative,shen2023proposal}, represented by the following Hamiltonian
\begin{equation}\label{tfi}
	\hat{H}_{\rm TFI}=h_{x}\sum_{j}\hat{X}_{j}+J\sum_{j}\hat{Z}_{j} \hat{Z}_{j+1}+i\gamma \sum_{j}\hat{Z}_{j},
\end{equation}
with Pauli operators $\hat{X}$ and $\hat{Z}$. Here, $h_x$ represents the transverse field strength, and $J$ quantifies the interaction strength between nearest-neighbor qubits. The term $i\gamma\hat{Z}_{j}$ represents an imaginary field  \cite{shen2023proposal}, that drives dynamics toward eigenstates with dominant imaginary components. In this scenario, observables, such as local magnetization, relax toward steady-state values over time, a characteristic behavior that our simulations are designed to capture.

To simulate nonunitary dynamics, we embed the desired 
$N$-qubit nonunitary operation $e^{-itH_{\rm TFI}}$ into a larger unitary circuit with an ancilla qubit (see Appendix.~\ref{apsec1}).  In this setup, measurement qubits represent the primary system from which observables are obtained, and the ancilla qubit is employed to filter results into the required computational sector. Post-selection on the ancilla $\uparrow$ is then used to extract the final result. In our work, the initial state is defined as $\ket{\psi_{0}}=\ket{\psi_{p}}\ket{\psi_{a}}$, composed of a spin-$\downarrow$ $\ket{\psi_{p}}=\ket{\downarrow\downarrow\downarrow\downarrow}$ initial state for measurement qubits, and  a spin-$\uparrow$ ancilla state $\ket{\psi_{a}}=\ket{\uparrow}$.   The simulation of this process is performed using variational circuits, optimized by minimizing the cost function [Eq.\ref{c}]. After post-selection, we can collect the final result from measurement qubits normalized as $e^{-itH_{\rm TFI}}\ket{\psi}/\left\| e^{-itH_{\rm TFI}}\ket{\psi}\right\|$. Here, $\ket{\psi}$ is for the state obtained for measurement qubits, and the outcome on the ancilla qubit is discarded after post-selection. Further technical details of implementation and optimization procedure are provided in Appendix.~\ref{apvqa}.

\rz{Importantly, the error-mitigation framework presented here is not restricted to simulations of nonunitary dynamics \cite{,qin2025dynamical,shen2024enhanced,yang2024non,xue2024topologically,gupta2024probabilistic}. Rather, it is broadly applicable across a wide range of quantum simulations, as the only essential requirement of our method is the construction of an identity-equivalent circuit that preserves the original circuit structure. This minimal condition allows for flexible integration into variational algorithms and other parameterized frameworks.} \rzz{We emphasize that, although extending our method to very large quantum circuits remains challenging, it shows strong potential for tasks that require repeated simulations of small-scale circuits with fixed structure. Such scenarios are particularly relevant in applications where robustness across a large number of circuit executions is more critical than circuit size. In the following section, we highlight potential applications of our approach to practical problems.}

\subsection{Benchmarking simulations}
\begin{figure}
	\centering
	\includegraphics[width=1\linewidth]{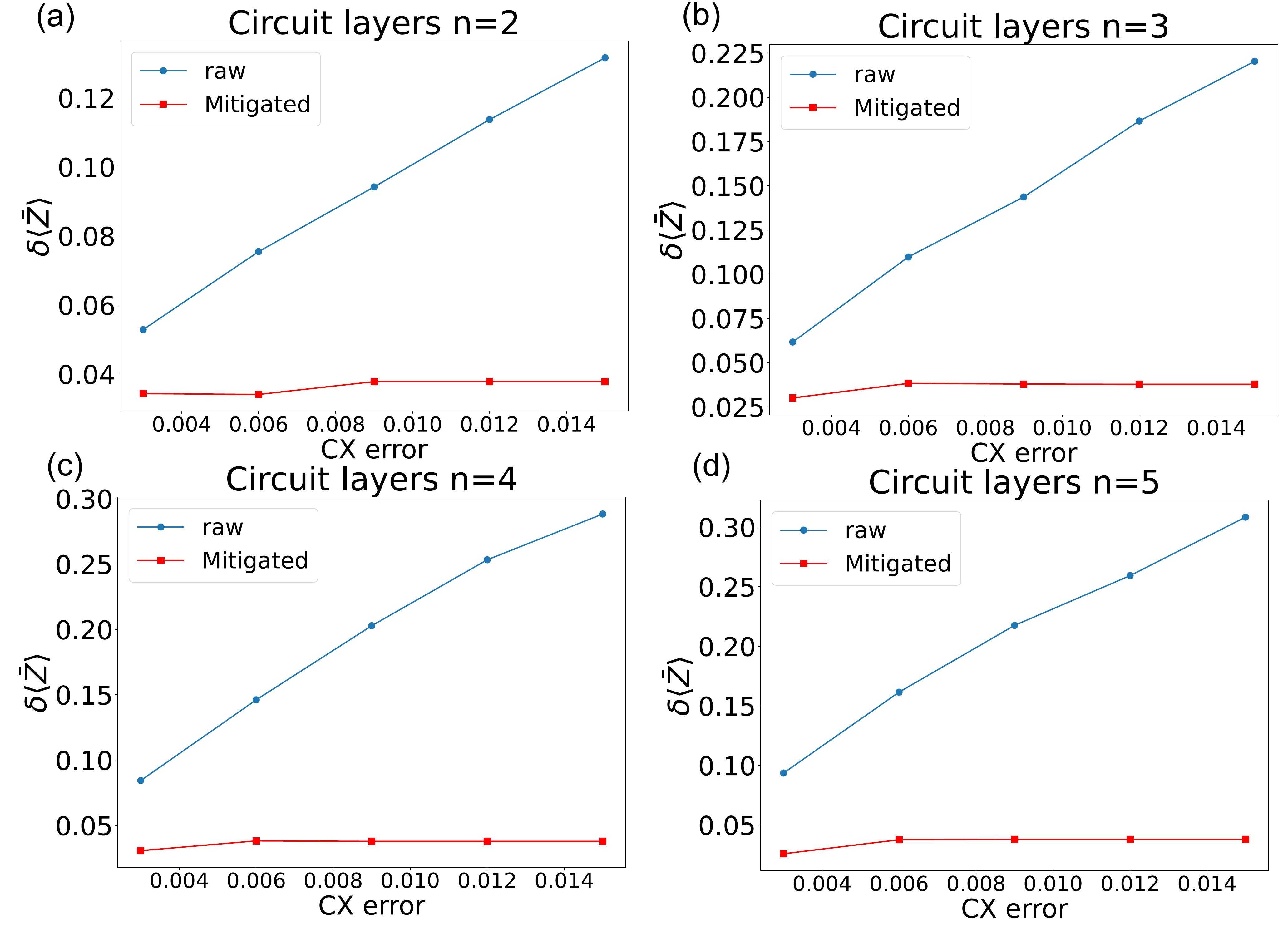}
	\caption{Demonstration of error reduction achieved by our structure-preserving mitigation method.
 We perform local noisy simulations of the time evolution $e^{-it\hat{H}_{\rm TFI}}$ governed by  Eq.\eqref{tfi} under increasing CX gate error rates (ranging from $0.003$ to $0.015$). Our error mitigation method follows the process described by Eq.~\ref{mit}. The effectiveness of our method is quantified by the average deviation defined in Eq.\ref{sz} measured under different circuit layers $n$. The red curve represents the mitigated result, showing consistently low deviation across all tested error rates. In contrast, the unmitigated results (blue curves) exhibit significant error accumulation. The initial state for measurement qubits is $\ket{\psi_{p}}=\ket{\downarrow\downarrow\downarrow\downarrow}$, and the ancilla state is $\ket{\psi_{a}}=\ket{\uparrow}$. $n$ is the number of circuit layers, and the structure of circuit layers is shown in FIG.\ref{fig:error}.  Other parameters are $J=1$, $h_{x}=1.5$ and $\gamma=-0.5$. The error rate of measurements is $1\%$.}
	\label{fig:zit}
\end{figure}
To validate the effectiveness of our enhanced error-mitigation strategy, illustrated in FIG.~\ref{fig:error},  we conduct classical simulations using a local noise simulator provided by the Qiskit framework \cite{Qiskit}.  This simulator allows precise control over error rates, providing a well-defined and controlled environment for assessing the performance of our method.

In noisy circuits, errors can arise from various sources, including readout errors \cite{yang2022efficient} and gate errors \cite{preskill2018quantum}. Readout errors can result in inaccurate measurement outcomes, that a target state ``1" might be incorrectly recorded as ``0" \cite{nachman2020unfolding,nation2021scalable}. Gate errors, on the other hand, can cause unintended operations, such as spin flips, with two-qubit gates, like CX gates, generally being the dominant contributors to noise. To isolate the impact of two-qubit gate errors in noisy classical simulations, we fix the measurement error rate and set the single-qubit gate error to zero. We then evaluate the performance of our error-mitigation method across varying levels of CX gate errors.

\begin{figure*}
    \centering
    \includegraphics[width=0.8\linewidth]{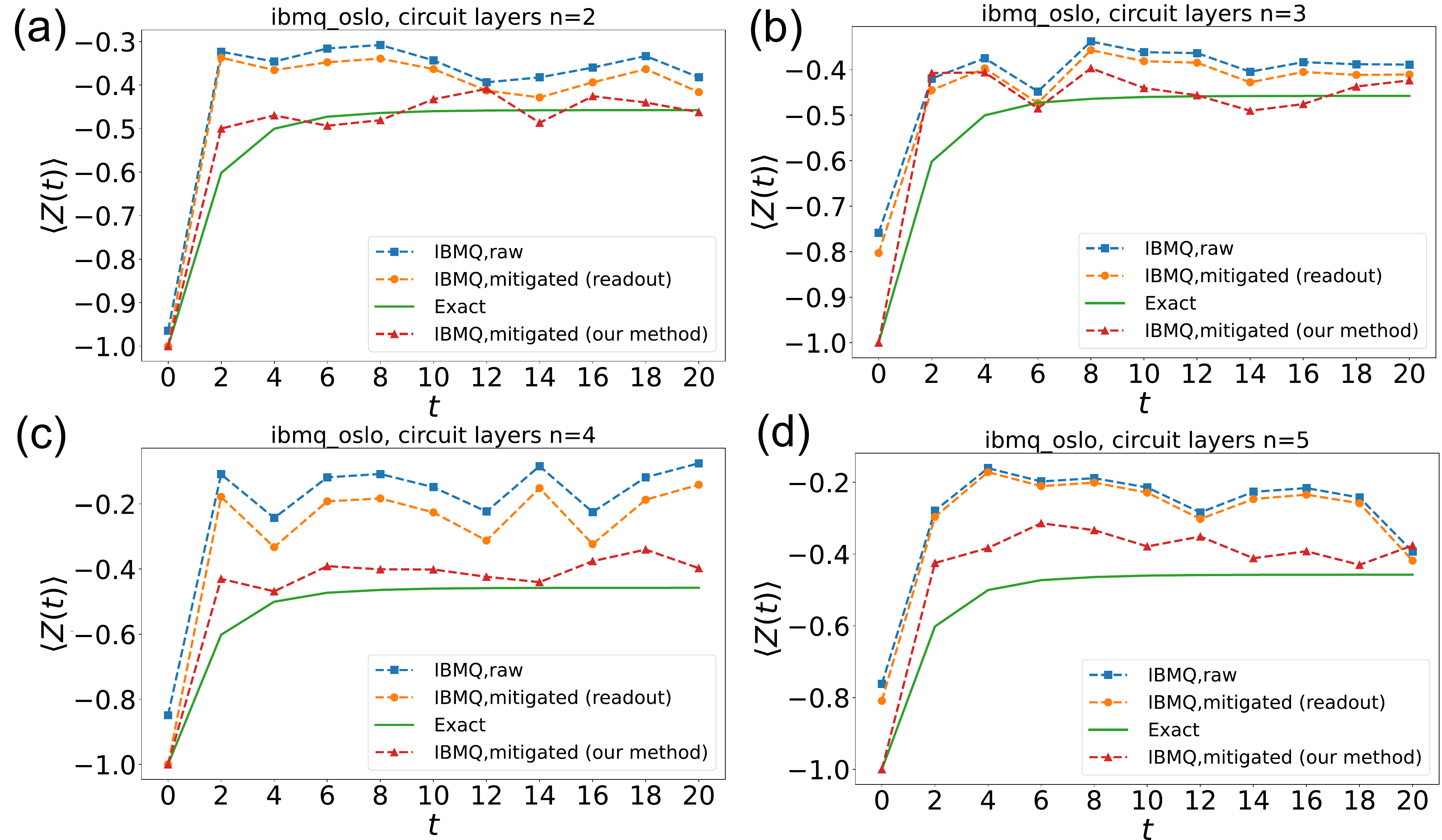}
    \caption{Comparison of error reduction: readout error mitigation vs. our structure-preserving error mitigation. In (a)-(d), we simulate the dynamics $e^{-it\hat{H}_{\rm TFI}}$ governed by the Ising chain [Eq.\eqref{tfi}] under increasing circuit layers $n$. For each depth, we compare the evolution of $Z$ magnetization between exact results (green curves) and noisy quantum simulations. The blue curve represents unmitigated results. The comparison between the red (out method) and yellow curves (readout mitigation) illustrates the significant improvement achieved through our approach. The initial state is set to $\ket{\psi{0}}=\ket{\psi_{p}}\ket{\psi_{a}}$ on a chain with 5 qubits. The initial state for measurement qubits is $\ket{\psi_{p}}=\ket{\downarrow\downarrow\downarrow\downarrow}$, and the ancilla state is $\ket{\psi_{a}}=\ket{\uparrow}$. Other parameters are $J=1$, $h_{x}=1.5$, and $\gamma=-0.5$. The number of shots is $32000$. The IBM Q device is ``ibmq Oslo" with a CX error rate of $0.012$ (see Appendix.~\ref{ibmq}).}
    \label{fig:read}
\end{figure*}

From outcomes, we can directly measure the average $Z$ magnetization, defined as
\begin{equation}\label{zz}
\langle Z(t)\rangle=\sum_{i}\bra{\psi(t)}Z_{i}\ket{\psi(t)}/N,
\end{equation}
where $\ket{\psi(t)}=e^{-it\hat{H}_{\rm TFI}}\ket{\psi(0)}/||e^{-it\hat{H}_{\rm TFI}}\ket{\psi(0)}||$, and $N$ is the number of measurement qubits. To quantify the impact of noise, we first compute the deviation between the exact and simulated results, expressed as follows:
\begin{equation}\label{z}
\delta\langle Z(t)\rangle=\langle Z(t)\rangle^{\rm Exact}-\langle Z(t)\rangle^{\rm  Mit/Raw},
\end{equation}
where $\langle Z(t)\rangle^{\rm Mit/Raw}$ denotes the simulated magnetization with (Mit) and without (Raw) error mitigation, respectively. To assess the overall robustness of simulations against noise, we calculate the following time-averaged deviation:
\begin{equation}\label{sz}
\delta\langle \bar{Z}\rangle=\sum^{T}_{k=0}\delta\langle Z(k\delta t)\rangle/T
\end{equation}
where $T=11$ is the total number of time steps, and we set $\delta t=2$. \rz{Note that these time steps refer to the evolution time of dynamics, and are different from circuit depth or layers.} 

Here, we first demonstrate our method in circuits built by $2$ to $5$ layers. As demonstrated in FIG.~\ref{fig:zit}, we measure the above deviation, averaged over the evolution $e^{-itH_{\rm TFI}}\ket{\psi}/\left\| e^{-itH_{\rm TFI}}\ket{\psi}\right\|$, with time $t$ evolving up to $20$. Here, our error-mitigated simulations exhibit consistently low deviation across all tested CX gate error rates and circuit layers. In contrast, the unmitigated results show a clear increase in deviation as the circuit depth grows.

The result above highlights the effectiveness of our error-mitigation approach. However, a key practical challenge in real-world quantum simulations lies in the inherent fluctuation of noise levels on physical quantum hardware. Current quantum processors often experience time-dependent variations in noise, making it difficult to maintain consistent noise profiles between the mitigation circuit $V^{\rm mit}$ and simulation circuit $V$. These inconsistencies can lead to inaccuracies in noise characterization. In our quantum hardware simulations, both mitigation and variational circuits are trained in advance. During execution, these circuits are combined and submitted to the IBM Quantum cloud. Although the same qubits are used throughout, fluctuations in noise levels remain unavoidable. Nevertheless, our subsequent simulations achieved on real hardware show that our error-mitigation strategy remains effective as long as noise fluctuations are not excessively large.

\subsection{Quantum hardware simulations}\label{sec4}

Before presenting the simulation results on IBM Quantum devices, we first provide a practical overview of the noise characteristics of quantum hardware. In this work, simulations are performed on an earlier-generation IBM Q device, which exhibits significantly higher noise levels than current devices. This allows for a clearer demonstration of the effectiveness of our proposed mitigation strategy.  According to the noise data from devices such as {\bf Ibmq\_oslo} shown in Appendix~\ref{ibmq}, the dominant source of noise in our circuits comes from stacked CX gates. In the following section, we demonstrate that our structure-preserving error mitigation method is capable of effectively suppressing these elusive gate errors on real quantum hardware.

In this section, we present our error-mitigated simulation on IBM Q devices. As previously discussed, readout errors caused by imperfect measurements are generally less significant than the error introduced by two-qubit gates. To explicitly assess their impact, we first isolate and analyze measurement noise by conducting simulations on the IBM Q device using only readout error mitigation.  In our implementation, readout error mitigation is performed using the {\bf mthree} package \cite{nation2021scalable}.

Results of quantum simulations performed on IBM Q devices, with and without readout error mitigation, are presented in FIG.\ref{fig:read}. Here, we simulate circuits with an increasing number of ansatz layers, ranging from 2 to 5, where deeper circuits correspond to greater noise accumulation.  In each case, we measure the average $Z$ magnetization, as shown in FIGs.\ref{fig:read} (a)-(d). As expected, discrepancies between exact results (green curves) and raw noisy simulations (blue curves) become more pronounced as the circuit depth increases. Although applying readout error mitigation (yellow curves) yields a slight improvement, such an effect is relatively insignificant. This limited enhancement indicates that the dominant source of error arises not from measurement imperfections but from the accumulation of noise in CX gates. In the following sections, we demonstrate that our structure-preserving error-mitigation method effectively addresses these gate errors, leading to a significant improvement in simulation fidelity.

Next, we perform quantum simulations using our optimized error-mitigation method. To implement this strategy, we first simulate $2^{4+1}$ variational identity circuits, each initialized with a distinct product state spanning the entire computational basis, in order to construct the corresponding calibration matrix.  Both the calibration process and the subsequent simulations are carried out within a short timeframe, thereby minimizing the influence of temporal fluctuations in hardware conditions. Consequently, it is reasonable to assume that any variations in device conditions during this period can be negligible.

Building on this, we compare the simulation results obtained using conventional readout error mitigation with those achieved through our enhanced structure-preserving method, as shown in FIGs.~\ref{fig:read} (a)-(d). For circuits with 2 to 4 layers, the mitigated result achieved through our method (red curves) shows excellent agreement with exact results (green curves), clearly demonstrating greater error reduction compared to the result (yellow curves) with only readout error mitigation. Notably, for the deepest circuit with five layers (FIG.~\ref{fig:read} (d)), the improvement of our approach is also obvious.

To further demonstrate the effectiveness of our method, we conduct additional simulations on three IBM Q devices with distinct noise profiles (see Appendix.~\ref{ibmq} for details). As shown in FIG.~\ref{fig:improve}, the unmitigated result (blue curves) shows a clear increase in error as the number of circuit layers increases, which is similar to the result shown in FIG.~\ref{fig:read}. However, the robustness of our method is evident.  The red curves, corresponding to the results mitigated by our circuit structure-preserving technique, maintain a stable and low error level of approximately $\langle \bar{Z}\rangle\approx 0.05$.  While our method significantly reduces the impact of gate errors, it does not eliminate noise effects. One limiting factor is the presence of sampling errors during the construction of the calibration matrix, which arises from the finite number of circuit executions. As a result, the matrix elements cannot be estimated with perfect precision, which is a practical limitation inherent in current hardware.

\begin{figure*}
    \centering
    \includegraphics[width=0.9\linewidth]{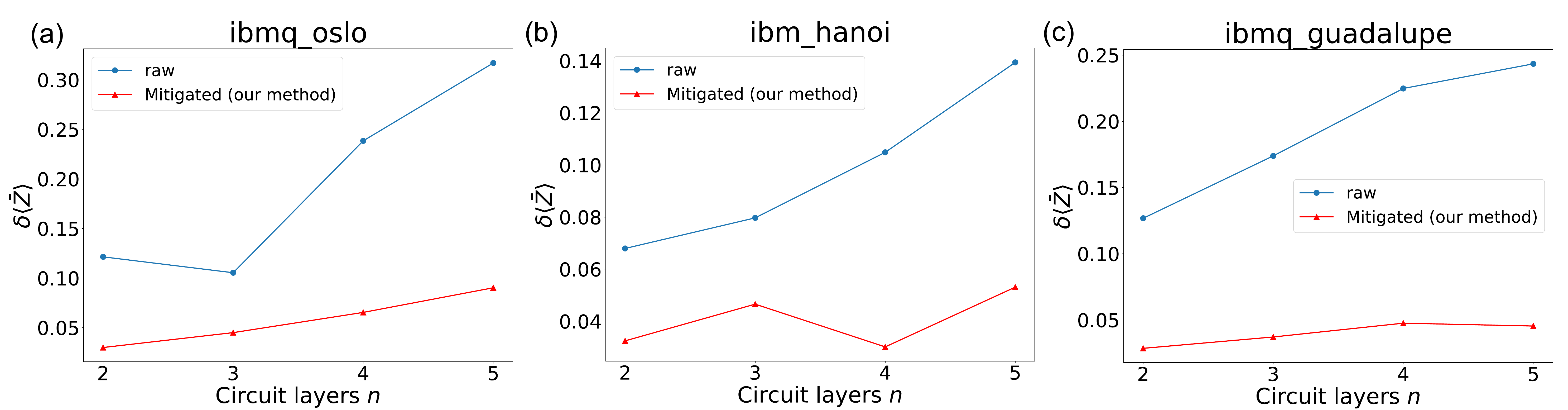}
    \caption{Demonstrations of error reduction via structure-preserving error mitigation on different devices.
    We present additional quantum simulations performed on three IBM Quantum devices, further evaluating the effectiveness of our structure-preserving method  [Eq.\ref{mitres}]. The performance is assessed using the average deviation defined in Eq.\ref{sz}. $n$ denotes the number of circuit layers.  For both devices, the red curves represent the error-mitigated results obtained using our method, exhibiting substantial error suppression across all tested circuit depths, compared to unmitigated results (blue curves).
    The initial state is set to $\ket{\psi{0}}=\ket{\psi_{p}}\ket{\psi_{a}}$ on a chain with 5 qubits. The initial state for measurement qubits is $\ket{\psi_{p}}=\ket{\downarrow\downarrow\downarrow\downarrow}$, and the ancilla state is $\ket{\psi_{a}}=\ket{\uparrow}$. Other parameters are $J=1$, $h_{x}=1.5$, and $\gamma=-0.5$. The number of shots is $32000$. Circuits are simulated on three IBM Q devices: ``ibmq Oslo" for (a), ``ibmq hanoi" for (b), and ``ibmq guadalupe" for (c), with CX errors of $0.012$ and $0.015$ respectively (see Appendix.~\ref{ibmq}).}
    \label{fig:improve}
\end{figure*}

\section{Conclusion}\label{sec5}
In this work, we present a circuit structure-preserving error mitigation framework designed to achieve high-fidelity quantum simulations on  NISQ devices. By leveraging on this approach, we achieve remarkably robust results on current noisy quantum hardware, significantly outperforming conventional readout-error mitigation. Even under increased qubit decoherence caused by deeper circuit structures, our method consistently maintains high accuracy across various circuit depths. This capability to effectively mitigate gate errors significantly extends the capability of current quantum hardware and opens new avenues for exploring exotic quantum systems  \cite{liu2022methods,xu2024non,xu2023digital}. Moreover, our approach holds significant potential for variational simulations beyond quantum many-body systems, including applications in molecular systems \cite{lanyon2010towards,wecker2014gate,kassal2011simulating,o2019calculating,gaita2019molecular} and material science \cite{ma2020quantum,bauer2020quantum,de2021materials}.

Our enhanced method is particularly well-suited for efficiently managing large-scale tasks involving thousands of small quantum circuits. Unlike the ZNE approach, which requires each circuit to be executed with an extended number of circuit layers, our approach significantly reduces both computational and quantum resource demands. By constructing a single calibration matrix that can be universally applied across all circuits, our method effectively minimizes quantum cost while maintaining robust error mitigation.

Furthermore, Quantum machine learning (QML) provides one of the most compelling application areas in quantum computing \cite{schuld2014quest,jeswal2019recent,altaisky2001quantum,ricks2003training}. Central to QML are parameterized quantum circuits, often referred to as quantum neural networks (QNNs), whose trainable gate parameters are iteratively optimized to minimize a task-specific cost function. These practical deployment on quantum hardware is still  hindered by gate errors. The circuit structure–preserving error mitigation strategy introduced in this work is naturally suited to overcoming these challenges. In QNN training, the circuit layout remains fixed across iterations, while only the gate parameters are updated. This aligns directly with the assumptions underlying our framework, where noise channels are approximately parameter-independent as long as the circuit structure is preserved. Thus, a single calibration matrix constructed from identity-equivalent mitigation circuits can be reused across the entire training process, avoiding the need to recalibrate at every optimization step.

\rz{We would also like to emphasize that the current IBM Quantum platform includes several well-established error mitigation techniques \cite{Qiskit}, such as ZNE and PEC. These frameworks are primarily designed to work with the Estimator primitive, which is used for computing expectation values of Pauli strings. However, our method is specifically developed for the Sampler primitive, which focuses on generating measurement outcome distributions. As such, our approach fills an important gap by enabling effective error mitigation in sampler simulations.}

\section*{ACKNOWLEDGMENTS}
We acknowledge the use of IBM Quantum services for this work. The views expressed are those of the authors and do not reflect the official policy or position of IBM or the IBM Quantum team. All data and codes of this work are available from the corresponding authors upon reasonable request. This work is supported by the Singapore Ministry of Education Academic Research Fund Tier-I preparatory grant (WBS no. A-8002656-00-00).

\newpage
\bibliography{ref}

\begin{thebibliography}{102}%
\makeatletter
\providecommand \@ifxundefined [1]{%
 \@ifx{#1\undefined}
}%
\providecommand \@ifnum [1]{%
 \ifnum #1\expandafter \@firstoftwo
 \else \expandafter \@secondoftwo
 \fi
}%
\providecommand \@ifx [1]{%
 \ifx #1\expandafter \@firstoftwo
 \else \expandafter \@secondoftwo
 \fi
}%
\providecommand \natexlab [1]{#1}%
\providecommand \enquote  [1]{``#1''}%
\providecommand \bibnamefont  [1]{#1}%
\providecommand \bibfnamefont [1]{#1}%
\providecommand \citenamefont [1]{#1}%
\providecommand \href@noop [0]{\@secondoftwo}%
\providecommand \href [0]{\begingroup \@sanitize@url \@href}%
\providecommand \@href[1]{\@@startlink{#1}\@@href}%
\providecommand \@@href[1]{\endgroup#1\@@endlink}%
\providecommand \@sanitize@url [0]{\catcode `\\12\catcode `\$12\catcode
  `\&12\catcode `\#12\catcode `\^12\catcode `\_12\catcode `\%12\relax}%
\providecommand \@@startlink[1]{}%
\providecommand \@@endlink[0]{}%
\providecommand \url  [0]{\begingroup\@sanitize@url \@url }%
\providecommand \@url [1]{\endgroup\@href {#1}{\urlprefix }}%
\providecommand \urlprefix  [0]{URL }%
\providecommand \Eprint [0]{\href }%
\providecommand \doibase [0]{https://doi.org/}%
\providecommand \selectlanguage [0]{\@gobble}%
\providecommand \bibinfo  [0]{\@secondoftwo}%
\providecommand \bibfield  [0]{\@secondoftwo}%
\providecommand \translation [1]{[#1]}%
\providecommand \BibitemOpen [0]{}%
\providecommand \bibitemStop [0]{}%
\providecommand \bibitemNoStop [0]{.\EOS\space}%
\providecommand \EOS [0]{\spacefactor3000\relax}%
\providecommand \BibitemShut  [1]{\csname bibitem#1\endcsname}%
\let\auto@bib@innerbib\@empty
\bibitem [{\citenamefont {Steane}(1998)}]{steane1998quantum}%
  \BibitemOpen
  \bibfield  {author} {\bibinfo {author} {\bibfnamefont {A.}~\bibnamefont
  {Steane}},\ }\bibfield  {title} {\bibinfo {title} {Quantum computing},\
  }\href@noop {} {\bibfield  {journal} {\bibinfo  {journal} {Reports on
  Progress in Physics}\ }\textbf {\bibinfo {volume} {61}},\ \bibinfo {pages}
  {117} (\bibinfo {year} {1998})}\BibitemShut {NoStop}%
\bibitem [{\citenamefont {Knill}(2010)}]{knill2010quantum}%
  \BibitemOpen
  \bibfield  {author} {\bibinfo {author} {\bibfnamefont {E.}~\bibnamefont
  {Knill}},\ }\bibfield  {title} {\bibinfo {title} {Quantum computing},\
  }\href@noop {} {\bibfield  {journal} {\bibinfo  {journal} {Nature}\ }\textbf
  {\bibinfo {volume} {463}},\ \bibinfo {pages} {441} (\bibinfo {year}
  {2010})}\BibitemShut {NoStop}%
\bibitem [{\citenamefont {Preskill}(2018)}]{preskill2018quantum}%
  \BibitemOpen
  \bibfield  {author} {\bibinfo {author} {\bibfnamefont {J.}~\bibnamefont
  {Preskill}},\ }\bibfield  {title} {\bibinfo {title} {Quantum computing in the
  nisq era and beyond},\ }\href@noop {} {\bibfield  {journal} {\bibinfo
  {journal} {Quantum}\ }\textbf {\bibinfo {volume} {2}},\ \bibinfo {pages} {79}
  (\bibinfo {year} {2018})}\BibitemShut {NoStop}%
\bibitem [{\citenamefont {National Academies~of Sciences}\ \emph
  {et~al.}(2019)\citenamefont {National Academies~of Sciences}, \citenamefont
  {Medicine} \emph {et~al.}}]{national2019quantum}%
  \BibitemOpen
  \bibfield  {author} {\bibinfo {author} {\bibfnamefont {E.}~\bibnamefont
  {National Academies~of Sciences}}, \bibinfo {author} {\bibnamefont
  {Medicine}}, \emph {et~al.},\ }\bibfield  {title} {\bibinfo {title} {Quantum
  computing: progress and prospects},\ }\href@noop {} {\  (\bibinfo {year}
  {2019})}\BibitemShut {NoStop}%
\bibitem [{\citenamefont {Randall}\ \emph {et~al.}(2021)\citenamefont
  {Randall}, \citenamefont {Bradley}, \citenamefont {van~der Gronden},
  \citenamefont {Galicia}, \citenamefont {Abobeih}, \citenamefont {Markham},
  \citenamefont {Twitchen}, \citenamefont {Machado}, \citenamefont {Yao},\ and\
  \citenamefont {Taminiau}}]{randall2021many}%
  \BibitemOpen
  \bibfield  {author} {\bibinfo {author} {\bibfnamefont {J.}~\bibnamefont
  {Randall}}, \bibinfo {author} {\bibfnamefont {C.}~\bibnamefont {Bradley}},
  \bibinfo {author} {\bibfnamefont {F.}~\bibnamefont {van~der Gronden}},
  \bibinfo {author} {\bibfnamefont {A.}~\bibnamefont {Galicia}}, \bibinfo
  {author} {\bibfnamefont {M.}~\bibnamefont {Abobeih}}, \bibinfo {author}
  {\bibfnamefont {M.}~\bibnamefont {Markham}}, \bibinfo {author} {\bibfnamefont
  {D.}~\bibnamefont {Twitchen}}, \bibinfo {author} {\bibfnamefont
  {F.}~\bibnamefont {Machado}}, \bibinfo {author} {\bibfnamefont
  {N.}~\bibnamefont {Yao}},\ and\ \bibinfo {author} {\bibfnamefont
  {T.}~\bibnamefont {Taminiau}},\ }\bibfield  {title} {\bibinfo {title}
  {Many-body--localized discrete time crystal with a programmable spin-based
  quantum simulator},\ }\href@noop {} {\bibfield  {journal} {\bibinfo
  {journal} {Science}\ }\textbf {\bibinfo {volume} {374}},\ \bibinfo {pages}
  {1474} (\bibinfo {year} {2021})}\BibitemShut {NoStop}%
\bibitem [{\citenamefont {Chen}\ \emph
  {et~al.}(2023{\natexlab{a}})\citenamefont {Chen}, \citenamefont {Shen},
  \citenamefont {Lee}, \citenamefont {Yang},\ and\ \citenamefont
  {Bomantara}}]{chen2023robust}%
  \BibitemOpen
  \bibfield  {author} {\bibinfo {author} {\bibfnamefont {T.}~\bibnamefont
  {Chen}}, \bibinfo {author} {\bibfnamefont {R.}~\bibnamefont {Shen}}, \bibinfo
  {author} {\bibfnamefont {C.~H.}\ \bibnamefont {Lee}}, \bibinfo {author}
  {\bibfnamefont {B.}~\bibnamefont {Yang}},\ and\ \bibinfo {author}
  {\bibfnamefont {R.~W.}\ \bibnamefont {Bomantara}},\ }\bibfield  {title}
  {\bibinfo {title} {A robust large-period discrete time crystal and its
  signature in a digital quantum computer},\ }\href@noop {} {\bibfield
  {journal} {\bibinfo  {journal} {arXiv preprint arXiv:2309.11560}\ } (\bibinfo
  {year} {2023}{\natexlab{a}})}\BibitemShut {NoStop}%
\bibitem [{\citenamefont {Zhou}\ \emph {et~al.}(2020)\citenamefont {Zhou},
  \citenamefont {Wang}, \citenamefont {Choi}, \citenamefont {Pichler},\ and\
  \citenamefont {Lukin}}]{zhou2020quantum}%
  \BibitemOpen
  \bibfield  {author} {\bibinfo {author} {\bibfnamefont {L.}~\bibnamefont
  {Zhou}}, \bibinfo {author} {\bibfnamefont {S.-T.}\ \bibnamefont {Wang}},
  \bibinfo {author} {\bibfnamefont {S.}~\bibnamefont {Choi}}, \bibinfo {author}
  {\bibfnamefont {H.}~\bibnamefont {Pichler}},\ and\ \bibinfo {author}
  {\bibfnamefont {M.~D.}\ \bibnamefont {Lukin}},\ }\bibfield  {title} {\bibinfo
  {title} {Quantum approximate optimization algorithm: Performance, mechanism,
  and implementation on near-term devices},\ }\href@noop {} {\bibfield
  {journal} {\bibinfo  {journal} {Physical Review X}\ }\textbf {\bibinfo
  {volume} {10}},\ \bibinfo {pages} {021067} (\bibinfo {year}
  {2020})}\BibitemShut {NoStop}%
\bibitem [{\citenamefont {Harrigan}\ \emph {et~al.}(2021)\citenamefont
  {Harrigan}, \citenamefont {Sung}, \citenamefont {Neeley}, \citenamefont
  {Satzinger}, \citenamefont {Arute}, \citenamefont {Arya}, \citenamefont
  {Atalaya}, \citenamefont {Bardin}, \citenamefont {Barends}, \citenamefont
  {Boixo} \emph {et~al.}}]{harrigan2021quantum}%
  \BibitemOpen
  \bibfield  {author} {\bibinfo {author} {\bibfnamefont {M.~P.}\ \bibnamefont
  {Harrigan}}, \bibinfo {author} {\bibfnamefont {K.~J.}\ \bibnamefont {Sung}},
  \bibinfo {author} {\bibfnamefont {M.}~\bibnamefont {Neeley}}, \bibinfo
  {author} {\bibfnamefont {K.~J.}\ \bibnamefont {Satzinger}}, \bibinfo {author}
  {\bibfnamefont {F.}~\bibnamefont {Arute}}, \bibinfo {author} {\bibfnamefont
  {K.}~\bibnamefont {Arya}}, \bibinfo {author} {\bibfnamefont {J.}~\bibnamefont
  {Atalaya}}, \bibinfo {author} {\bibfnamefont {J.~C.}\ \bibnamefont {Bardin}},
  \bibinfo {author} {\bibfnamefont {R.}~\bibnamefont {Barends}}, \bibinfo
  {author} {\bibfnamefont {S.}~\bibnamefont {Boixo}}, \emph {et~al.},\
  }\bibfield  {title} {\bibinfo {title} {Quantum approximate optimization of
  non-planar graph problems on a planar superconducting processor},\
  }\href@noop {} {\bibfield  {journal} {\bibinfo  {journal} {Nature Physics}\
  }\textbf {\bibinfo {volume} {17}},\ \bibinfo {pages} {332} (\bibinfo {year}
  {2021})}\BibitemShut {NoStop}%
\bibitem [{\citenamefont {Montanaro}\ and\ \citenamefont
  {Zhou}(2024)}]{montanaro2024quantum}%
  \BibitemOpen
  \bibfield  {author} {\bibinfo {author} {\bibfnamefont {A.}~\bibnamefont
  {Montanaro}}\ and\ \bibinfo {author} {\bibfnamefont {L.}~\bibnamefont
  {Zhou}},\ }\bibfield  {title} {\bibinfo {title} {Quantum speedups in solving
  near-symmetric optimization problems by low-depth qaoa},\ }\href@noop {}
  {\bibfield  {journal} {\bibinfo  {journal} {arXiv preprint arXiv:2411.04979}\
  } (\bibinfo {year} {2024})}\BibitemShut {NoStop}%
\bibitem [{\citenamefont {Koh}\ \emph {et~al.}(2024)\citenamefont {Koh},
  \citenamefont {Tai},\ and\ \citenamefont {Lee}}]{koh2024realization}%
  \BibitemOpen
  \bibfield  {author} {\bibinfo {author} {\bibfnamefont {J.~M.}\ \bibnamefont
  {Koh}}, \bibinfo {author} {\bibfnamefont {T.}~\bibnamefont {Tai}},\ and\
  \bibinfo {author} {\bibfnamefont {C.~H.}\ \bibnamefont {Lee}},\ }\bibfield
  {title} {\bibinfo {title} {Realization of higher-order topological lattices
  on a quantum computer},\ }\href@noop {} {\bibfield  {journal} {\bibinfo
  {journal} {Nature Communications}\ }\textbf {\bibinfo {volume} {15}},\
  \bibinfo {pages} {5807} (\bibinfo {year} {2024})}\BibitemShut {NoStop}%
\bibitem [{\citenamefont {Chen}\ \emph {et~al.}(2024)\citenamefont {Chen},
  \citenamefont {Ding}, \citenamefont {Shen}, \citenamefont {Zhu},\ and\
  \citenamefont {Gong}}]{chen2024direct}%
  \BibitemOpen
  \bibfield  {author} {\bibinfo {author} {\bibfnamefont {T.}~\bibnamefont
  {Chen}}, \bibinfo {author} {\bibfnamefont {H.-T.}\ \bibnamefont {Ding}},
  \bibinfo {author} {\bibfnamefont {R.}~\bibnamefont {Shen}}, \bibinfo {author}
  {\bibfnamefont {S.-L.}\ \bibnamefont {Zhu}},\ and\ \bibinfo {author}
  {\bibfnamefont {J.}~\bibnamefont {Gong}},\ }\bibfield  {title} {\bibinfo
  {title} {Direct probe of topology and geometry of quantum states on the ibm q
  quantum processor},\ }\href@noop {} {\bibfield  {journal} {\bibinfo
  {journal} {Physical Review B}\ }\textbf {\bibinfo {volume} {110}},\ \bibinfo
  {pages} {205402} (\bibinfo {year} {2024})}\BibitemShut {NoStop}%
\bibitem [{\citenamefont {Koh}\ \emph {et~al.}(2022{\natexlab{a}})\citenamefont
  {Koh}, \citenamefont {Tai},\ and\ \citenamefont {Lee}}]{koh2022simulation}%
  \BibitemOpen
  \bibfield  {author} {\bibinfo {author} {\bibfnamefont {J.~M.}\ \bibnamefont
  {Koh}}, \bibinfo {author} {\bibfnamefont {T.}~\bibnamefont {Tai}},\ and\
  \bibinfo {author} {\bibfnamefont {C.~H.}\ \bibnamefont {Lee}},\ }\bibfield
  {title} {\bibinfo {title} {Simulation of interaction-induced chiral
  topological dynamics on a digital quantum computer},\ }\href@noop {}
  {\bibfield  {journal} {\bibinfo  {journal} {Physical Review Letters}\
  }\textbf {\bibinfo {volume} {129}},\ \bibinfo {pages} {140502} (\bibinfo
  {year} {2022}{\natexlab{a}})}\BibitemShut {NoStop}%
\bibitem [{\citenamefont {Koh}\ \emph {et~al.}(2022{\natexlab{b}})\citenamefont
  {Koh}, \citenamefont {Tai}, \citenamefont {Phee}, \citenamefont {Ng},\ and\
  \citenamefont {Lee}}]{koh2022stabilizing}%
  \BibitemOpen
  \bibfield  {author} {\bibinfo {author} {\bibfnamefont {J.~M.}\ \bibnamefont
  {Koh}}, \bibinfo {author} {\bibfnamefont {T.}~\bibnamefont {Tai}}, \bibinfo
  {author} {\bibfnamefont {Y.~H.}\ \bibnamefont {Phee}}, \bibinfo {author}
  {\bibfnamefont {W.~E.}\ \bibnamefont {Ng}},\ and\ \bibinfo {author}
  {\bibfnamefont {C.~H.}\ \bibnamefont {Lee}},\ }\bibfield  {title} {\bibinfo
  {title} {Stabilizing multiple topological fermions on a quantum computer},\
  }\href@noop {} {\bibfield  {journal} {\bibinfo  {journal} {npj Quantum
  Information}\ }\textbf {\bibinfo {volume} {8}},\ \bibinfo {pages} {16}
  (\bibinfo {year} {2022}{\natexlab{b}})}\BibitemShut {NoStop}%
\bibitem [{\citenamefont {Ippoliti}\ \emph {et~al.}(2021)\citenamefont
  {Ippoliti}, \citenamefont {Kechedzhi}, \citenamefont {Moessner},
  \citenamefont {Sondhi},\ and\ \citenamefont {Khemani}}]{ippoliti2021many}%
  \BibitemOpen
  \bibfield  {author} {\bibinfo {author} {\bibfnamefont {M.}~\bibnamefont
  {Ippoliti}}, \bibinfo {author} {\bibfnamefont {K.}~\bibnamefont {Kechedzhi}},
  \bibinfo {author} {\bibfnamefont {R.}~\bibnamefont {Moessner}}, \bibinfo
  {author} {\bibfnamefont {S.}~\bibnamefont {Sondhi}},\ and\ \bibinfo {author}
  {\bibfnamefont {V.}~\bibnamefont {Khemani}},\ }\bibfield  {title} {\bibinfo
  {title} {Many-body physics in the nisq era: quantum programming a discrete
  time crystal},\ }\href@noop {} {\bibfield  {journal} {\bibinfo  {journal}
  {PRX Quantum}\ }\textbf {\bibinfo {volume} {2}},\ \bibinfo {pages} {030346}
  (\bibinfo {year} {2021})}\BibitemShut {NoStop}%
\bibitem [{\citenamefont {Xu}\ \emph {et~al.}(2021)\citenamefont {Xu},
  \citenamefont {Zhang}, \citenamefont {Han}, \citenamefont {Li}, \citenamefont
  {Xue}, \citenamefont {Liu}, \citenamefont {Jin},\ and\ \citenamefont
  {Yu}}]{xu2021realizing}%
  \BibitemOpen
  \bibfield  {author} {\bibinfo {author} {\bibfnamefont {H.}~\bibnamefont
  {Xu}}, \bibinfo {author} {\bibfnamefont {J.}~\bibnamefont {Zhang}}, \bibinfo
  {author} {\bibfnamefont {J.}~\bibnamefont {Han}}, \bibinfo {author}
  {\bibfnamefont {Z.}~\bibnamefont {Li}}, \bibinfo {author} {\bibfnamefont
  {G.}~\bibnamefont {Xue}}, \bibinfo {author} {\bibfnamefont {W.}~\bibnamefont
  {Liu}}, \bibinfo {author} {\bibfnamefont {Y.}~\bibnamefont {Jin}},\ and\
  \bibinfo {author} {\bibfnamefont {H.}~\bibnamefont {Yu}},\ }\bibfield
  {title} {\bibinfo {title} {Realizing discrete time crystal in an
  one-dimensional superconducting qubit chain},\ }\href@noop {} {\bibfield
  {journal} {\bibinfo  {journal} {arXiv preprint arXiv:2108.00942}\ } (\bibinfo
  {year} {2021})}\BibitemShut {NoStop}%
\bibitem [{\citenamefont {Brennen}\ and\ \citenamefont
  {Miyake}(2008)}]{brennen2008measurement}%
  \BibitemOpen
  \bibfield  {author} {\bibinfo {author} {\bibfnamefont {G.~K.}\ \bibnamefont
  {Brennen}}\ and\ \bibinfo {author} {\bibfnamefont {A.}~\bibnamefont
  {Miyake}},\ }\bibfield  {title} {\bibinfo {title} {Measurement-based quantum
  computer in the gapped ground state of a two-body hamiltonian},\ }\href@noop
  {} {\bibfield  {journal} {\bibinfo  {journal} {Physical review letters}\
  }\textbf {\bibinfo {volume} {101}},\ \bibinfo {pages} {010502} (\bibinfo
  {year} {2008})}\BibitemShut {NoStop}%
\bibitem [{\citenamefont {Poulin}\ and\ \citenamefont
  {Wocjan}(2009)}]{poulin2009preparing}%
  \BibitemOpen
  \bibfield  {author} {\bibinfo {author} {\bibfnamefont {D.}~\bibnamefont
  {Poulin}}\ and\ \bibinfo {author} {\bibfnamefont {P.}~\bibnamefont
  {Wocjan}},\ }\bibfield  {title} {\bibinfo {title} {Preparing ground states of
  quantum many-body systems on a quantum computer},\ }\href@noop {} {\bibfield
  {journal} {\bibinfo  {journal} {Physical review letters}\ }\textbf {\bibinfo
  {volume} {102}},\ \bibinfo {pages} {130503} (\bibinfo {year}
  {2009})}\BibitemShut {NoStop}%
\bibitem [{\citenamefont {Wecker}\ \emph {et~al.}(2015)\citenamefont {Wecker},
  \citenamefont {Hastings}, \citenamefont {Wiebe}, \citenamefont {Clark},
  \citenamefont {Nayak},\ and\ \citenamefont {Troyer}}]{wecker2015solving}%
  \BibitemOpen
  \bibfield  {author} {\bibinfo {author} {\bibfnamefont {D.}~\bibnamefont
  {Wecker}}, \bibinfo {author} {\bibfnamefont {M.~B.}\ \bibnamefont
  {Hastings}}, \bibinfo {author} {\bibfnamefont {N.}~\bibnamefont {Wiebe}},
  \bibinfo {author} {\bibfnamefont {B.~K.}\ \bibnamefont {Clark}}, \bibinfo
  {author} {\bibfnamefont {C.}~\bibnamefont {Nayak}},\ and\ \bibinfo {author}
  {\bibfnamefont {M.}~\bibnamefont {Troyer}},\ }\bibfield  {title} {\bibinfo
  {title} {Solving strongly correlated electron models on a quantum computer},\
  }\href@noop {} {\bibfield  {journal} {\bibinfo  {journal} {Physical Review
  A}\ }\textbf {\bibinfo {volume} {92}},\ \bibinfo {pages} {062318} (\bibinfo
  {year} {2015})}\BibitemShut {NoStop}%
\bibitem [{\citenamefont {Nam}\ \emph {et~al.}(2020)\citenamefont {Nam},
  \citenamefont {Chen}, \citenamefont {Pisenti}, \citenamefont {Wright},
  \citenamefont {Delaney}, \citenamefont {Maslov}, \citenamefont {Brown},
  \citenamefont {Allen}, \citenamefont {Amini}, \citenamefont {Apisdorf} \emph
  {et~al.}}]{nam2020ground}%
  \BibitemOpen
  \bibfield  {author} {\bibinfo {author} {\bibfnamefont {Y.}~\bibnamefont
  {Nam}}, \bibinfo {author} {\bibfnamefont {J.-S.}\ \bibnamefont {Chen}},
  \bibinfo {author} {\bibfnamefont {N.~C.}\ \bibnamefont {Pisenti}}, \bibinfo
  {author} {\bibfnamefont {K.}~\bibnamefont {Wright}}, \bibinfo {author}
  {\bibfnamefont {C.}~\bibnamefont {Delaney}}, \bibinfo {author} {\bibfnamefont
  {D.}~\bibnamefont {Maslov}}, \bibinfo {author} {\bibfnamefont {K.~R.}\
  \bibnamefont {Brown}}, \bibinfo {author} {\bibfnamefont {S.}~\bibnamefont
  {Allen}}, \bibinfo {author} {\bibfnamefont {J.~M.}\ \bibnamefont {Amini}},
  \bibinfo {author} {\bibfnamefont {J.}~\bibnamefont {Apisdorf}}, \emph
  {et~al.},\ }\bibfield  {title} {\bibinfo {title} {Ground-state energy
  estimation of the water molecule on a trapped-ion quantum computer},\
  }\href@noop {} {\bibfield  {journal} {\bibinfo  {journal} {npj Quantum
  Information}\ }\textbf {\bibinfo {volume} {6}},\ \bibinfo {pages} {33}
  (\bibinfo {year} {2020})}\BibitemShut {NoStop}%
\bibitem [{\citenamefont {Stanisic}\ \emph {et~al.}(2022)\citenamefont
  {Stanisic}, \citenamefont {Bosse}, \citenamefont {Gambetta}, \citenamefont
  {Santos}, \citenamefont {Mruczkiewicz}, \citenamefont {O’Brien},
  \citenamefont {Ostby},\ and\ \citenamefont
  {Montanaro}}]{stanisic2022observing}%
  \BibitemOpen
  \bibfield  {author} {\bibinfo {author} {\bibfnamefont {S.}~\bibnamefont
  {Stanisic}}, \bibinfo {author} {\bibfnamefont {J.~L.}\ \bibnamefont {Bosse}},
  \bibinfo {author} {\bibfnamefont {F.~M.}\ \bibnamefont {Gambetta}}, \bibinfo
  {author} {\bibfnamefont {R.~A.}\ \bibnamefont {Santos}}, \bibinfo {author}
  {\bibfnamefont {W.}~\bibnamefont {Mruczkiewicz}}, \bibinfo {author}
  {\bibfnamefont {T.~E.}\ \bibnamefont {O’Brien}}, \bibinfo {author}
  {\bibfnamefont {E.}~\bibnamefont {Ostby}},\ and\ \bibinfo {author}
  {\bibfnamefont {A.}~\bibnamefont {Montanaro}},\ }\bibfield  {title} {\bibinfo
  {title} {Observing ground-state properties of the fermi-hubbard model using a
  scalable algorithm on a quantum computer},\ }\href@noop {} {\bibfield
  {journal} {\bibinfo  {journal} {Nature Communications}\ }\textbf {\bibinfo
  {volume} {13}},\ \bibinfo {pages} {5743} (\bibinfo {year}
  {2022})}\BibitemShut {NoStop}%
\bibitem [{\citenamefont {Shen}\ \emph
  {et~al.}(2025{\natexlab{a}})\citenamefont {Shen}, \citenamefont {Chen},
  \citenamefont {Yang}, \citenamefont {Zhong},\ and\ \citenamefont
  {Lee}}]{shen2025robust}%
  \BibitemOpen
  \bibfield  {author} {\bibinfo {author} {\bibfnamefont {R.}~\bibnamefont
  {Shen}}, \bibinfo {author} {\bibfnamefont {T.}~\bibnamefont {Chen}}, \bibinfo
  {author} {\bibfnamefont {B.}~\bibnamefont {Yang}}, \bibinfo {author}
  {\bibfnamefont {Y.}~\bibnamefont {Zhong}},\ and\ \bibinfo {author}
  {\bibfnamefont {C.~H.}\ \bibnamefont {Lee}},\ }\bibfield  {title} {\bibinfo
  {title} {Robust simulations of many-body symmetry-protected topological phase
  transitions on a quantum processor},\ }\href@noop {} {\bibfield  {journal}
  {\bibinfo  {journal} {arXiv preprint arXiv:2503.08776}\ } (\bibinfo {year}
  {2025}{\natexlab{a}})}\BibitemShut {NoStop}%
\bibitem [{\citenamefont {Sciorilli}\ \emph {et~al.}(2025)\citenamefont
  {Sciorilli}, \citenamefont {Borges}, \citenamefont {Patti}, \citenamefont
  {Garc{\'\i}a-Mart{\'\i}n}, \citenamefont {Camilo}, \citenamefont
  {Anandkumar},\ and\ \citenamefont {Aolita}}]{sciorilli2025towards}%
  \BibitemOpen
  \bibfield  {author} {\bibinfo {author} {\bibfnamefont {M.}~\bibnamefont
  {Sciorilli}}, \bibinfo {author} {\bibfnamefont {L.}~\bibnamefont {Borges}},
  \bibinfo {author} {\bibfnamefont {T.~L.}\ \bibnamefont {Patti}}, \bibinfo
  {author} {\bibfnamefont {D.}~\bibnamefont {Garc{\'\i}a-Mart{\'\i}n}},
  \bibinfo {author} {\bibfnamefont {G.}~\bibnamefont {Camilo}}, \bibinfo
  {author} {\bibfnamefont {A.}~\bibnamefont {Anandkumar}},\ and\ \bibinfo
  {author} {\bibfnamefont {L.}~\bibnamefont {Aolita}},\ }\bibfield  {title}
  {\bibinfo {title} {Towards large-scale quantum optimization solvers with few
  qubits},\ }\href@noop {} {\bibfield  {journal} {\bibinfo  {journal} {Nature
  Communications}\ }\textbf {\bibinfo {volume} {16}},\ \bibinfo {pages} {476}
  (\bibinfo {year} {2025})}\BibitemShut {NoStop}%
\bibitem [{\citenamefont {Kirmani}\ \emph {et~al.}(2022)\citenamefont
  {Kirmani}, \citenamefont {Bull}, \citenamefont {Hou}, \citenamefont
  {Saravanan}, \citenamefont {Saeed}, \citenamefont {Papi{\'c}}, \citenamefont
  {Rahmani},\ and\ \citenamefont {Ghaemi}}]{kirmani2022probing}%
  \BibitemOpen
  \bibfield  {author} {\bibinfo {author} {\bibfnamefont {A.}~\bibnamefont
  {Kirmani}}, \bibinfo {author} {\bibfnamefont {K.}~\bibnamefont {Bull}},
  \bibinfo {author} {\bibfnamefont {C.-Y.}\ \bibnamefont {Hou}}, \bibinfo
  {author} {\bibfnamefont {V.}~\bibnamefont {Saravanan}}, \bibinfo {author}
  {\bibfnamefont {S.~M.}\ \bibnamefont {Saeed}}, \bibinfo {author}
  {\bibfnamefont {Z.}~\bibnamefont {Papi{\'c}}}, \bibinfo {author}
  {\bibfnamefont {A.}~\bibnamefont {Rahmani}},\ and\ \bibinfo {author}
  {\bibfnamefont {P.}~\bibnamefont {Ghaemi}},\ }\bibfield  {title} {\bibinfo
  {title} {Probing geometric excitations of fractional quantum hall states on
  quantum computers},\ }\href@noop {} {\bibfield  {journal} {\bibinfo
  {journal} {Physical Review Letters}\ }\textbf {\bibinfo {volume} {129}},\
  \bibinfo {pages} {056801} (\bibinfo {year} {2022})}\BibitemShut {NoStop}%
\bibitem [{\citenamefont {Agresti}\ \emph {et~al.}(2024)\citenamefont
  {Agresti}, \citenamefont {Paul}, \citenamefont {Schiansky}, \citenamefont
  {Steiner}, \citenamefont {Yin}, \citenamefont {Pentangelo}, \citenamefont
  {Piacentini}, \citenamefont {Crespi}, \citenamefont {Ban}, \citenamefont
  {Ceccarelli} \emph {et~al.}}]{agresti2024demonstration}%
  \BibitemOpen
  \bibfield  {author} {\bibinfo {author} {\bibfnamefont {I.}~\bibnamefont
  {Agresti}}, \bibinfo {author} {\bibfnamefont {K.}~\bibnamefont {Paul}},
  \bibinfo {author} {\bibfnamefont {P.}~\bibnamefont {Schiansky}}, \bibinfo
  {author} {\bibfnamefont {S.}~\bibnamefont {Steiner}}, \bibinfo {author}
  {\bibfnamefont {Z.}~\bibnamefont {Yin}}, \bibinfo {author} {\bibfnamefont
  {C.}~\bibnamefont {Pentangelo}}, \bibinfo {author} {\bibfnamefont
  {S.}~\bibnamefont {Piacentini}}, \bibinfo {author} {\bibfnamefont
  {A.}~\bibnamefont {Crespi}}, \bibinfo {author} {\bibfnamefont
  {Y.}~\bibnamefont {Ban}}, \bibinfo {author} {\bibfnamefont {F.}~\bibnamefont
  {Ceccarelli}}, \emph {et~al.},\ }\bibfield  {title} {\bibinfo {title}
  {Demonstration of hardware efficient photonic variational quantum
  algorithm},\ }\href@noop {} {\bibfield  {journal} {\bibinfo  {journal} {arXiv
  preprint arXiv:2408.10339}\ } (\bibinfo {year} {2024})}\BibitemShut {NoStop}%
\bibitem [{\citenamefont {Wang}\ \emph {et~al.}(2024)\citenamefont {Wang},
  \citenamefont {Ding}, \citenamefont {C{\'a}rdenas-L{\'o}pez},\ and\
  \citenamefont {Chen}}]{wang2024pulse}%
  \BibitemOpen
  \bibfield  {author} {\bibinfo {author} {\bibfnamefont {Y.}~\bibnamefont
  {Wang}}, \bibinfo {author} {\bibfnamefont {Y.}~\bibnamefont {Ding}}, \bibinfo
  {author} {\bibfnamefont {F.~A.}\ \bibnamefont {C{\'a}rdenas-L{\'o}pez}},\
  and\ \bibinfo {author} {\bibfnamefont {X.}~\bibnamefont {Chen}},\ }\bibfield
  {title} {\bibinfo {title} {Pulse-based variational quantum optimization and
  metalearning in superconducting circuits},\ }\href@noop {} {\bibfield
  {journal} {\bibinfo  {journal} {Physical Review Applied}\ }\textbf {\bibinfo
  {volume} {22}},\ \bibinfo {pages} {024009} (\bibinfo {year}
  {2024})}\BibitemShut {NoStop}%
\bibitem [{\citenamefont {Koh}\ \emph {et~al.}(2025)\citenamefont {Koh},
  \citenamefont {Xue}, \citenamefont {Tai}, \citenamefont {Koh},\ and\
  \citenamefont {Lee}}]{koh2025interacting}%
  \BibitemOpen
  \bibfield  {author} {\bibinfo {author} {\bibfnamefont {J.~M.}\ \bibnamefont
  {Koh}}, \bibinfo {author} {\bibfnamefont {W.-T.}\ \bibnamefont {Xue}},
  \bibinfo {author} {\bibfnamefont {T.}~\bibnamefont {Tai}}, \bibinfo {author}
  {\bibfnamefont {D.~E.}\ \bibnamefont {Koh}},\ and\ \bibinfo {author}
  {\bibfnamefont {C.~H.}\ \bibnamefont {Lee}},\ }\bibfield  {title} {\bibinfo
  {title} {Interacting non-hermitian edge and cluster bursts on a digital
  quantum processor},\ }\href@noop {} {\bibfield  {journal} {\bibinfo
  {journal} {arXiv preprint arXiv:2503.14595}\ } (\bibinfo {year}
  {2025})}\BibitemShut {NoStop}%
\bibitem [{\citenamefont {Gingrich}\ and\ \citenamefont
  {Williams}(2004)}]{gingrich2004non}%
  \BibitemOpen
  \bibfield  {author} {\bibinfo {author} {\bibfnamefont {R.~M.}\ \bibnamefont
  {Gingrich}}\ and\ \bibinfo {author} {\bibfnamefont {C.~P.}\ \bibnamefont
  {Williams}},\ }\bibfield  {title} {\bibinfo {title} {Non-unitary
  probabilistic quantum computing},\ }\href@noop {} {\  (\bibinfo {year}
  {2004})}\BibitemShut {NoStop}%
\bibitem [{\citenamefont {Williams}(2004)}]{williams2004probabilistic}%
  \BibitemOpen
  \bibfield  {author} {\bibinfo {author} {\bibfnamefont {C.~P.}\ \bibnamefont
  {Williams}},\ }\bibfield  {title} {\bibinfo {title} {Probabilistic nonunitary
  quantum computing},\ }in\ \href@noop {} {\emph {\bibinfo {booktitle} {Quantum
  Information and Computation II}}},\ Vol.\ \bibinfo {volume} {5436}\ (\bibinfo
  {organization} {SPIE},\ \bibinfo {year} {2004})\ pp.\ \bibinfo {pages}
  {297--306}\BibitemShut {NoStop}%
\bibitem [{\citenamefont {Hu}\ \emph {et~al.}(2020)\citenamefont {Hu},
  \citenamefont {Xia},\ and\ \citenamefont {Kais}}]{hu2020quantum}%
  \BibitemOpen
  \bibfield  {author} {\bibinfo {author} {\bibfnamefont {Z.}~\bibnamefont
  {Hu}}, \bibinfo {author} {\bibfnamefont {R.}~\bibnamefont {Xia}},\ and\
  \bibinfo {author} {\bibfnamefont {S.}~\bibnamefont {Kais}},\ }\bibfield
  {title} {\bibinfo {title} {A quantum algorithm for evolving open quantum
  dynamics on quantum computing devices},\ }\href@noop {} {\bibfield  {journal}
  {\bibinfo  {journal} {Scientific reports}\ }\textbf {\bibinfo {volume}
  {10}},\ \bibinfo {pages} {3301} (\bibinfo {year} {2020})}\BibitemShut
  {NoStop}%
\bibitem [{\citenamefont {Head-Marsden}\ \emph {et~al.}(2021)\citenamefont
  {Head-Marsden}, \citenamefont {Krastanov}, \citenamefont {Mazziotti},\ and\
  \citenamefont {Narang}}]{head2021capturing}%
  \BibitemOpen
  \bibfield  {author} {\bibinfo {author} {\bibfnamefont {K.}~\bibnamefont
  {Head-Marsden}}, \bibinfo {author} {\bibfnamefont {S.}~\bibnamefont
  {Krastanov}}, \bibinfo {author} {\bibfnamefont {D.~A.}\ \bibnamefont
  {Mazziotti}},\ and\ \bibinfo {author} {\bibfnamefont {P.}~\bibnamefont
  {Narang}},\ }\bibfield  {title} {\bibinfo {title} {Capturing non-markovian
  dynamics on near-term quantum computers},\ }\href@noop {} {\bibfield
  {journal} {\bibinfo  {journal} {Physical Review Research}\ }\textbf {\bibinfo
  {volume} {3}},\ \bibinfo {pages} {013182} (\bibinfo {year}
  {2021})}\BibitemShut {NoStop}%
\bibitem [{\citenamefont {Cai}\ \emph {et~al.}(2023)\citenamefont {Cai},
  \citenamefont {Babbush}, \citenamefont {Benjamin}, \citenamefont {Endo},
  \citenamefont {Huggins}, \citenamefont {Li}, \citenamefont {McClean},\ and\
  \citenamefont {O'Brien}}]{RevModPhys.95.045005}%
  \BibitemOpen
  \bibfield  {author} {\bibinfo {author} {\bibfnamefont {Z.}~\bibnamefont
  {Cai}}, \bibinfo {author} {\bibfnamefont {R.}~\bibnamefont {Babbush}},
  \bibinfo {author} {\bibfnamefont {S.~C.}\ \bibnamefont {Benjamin}}, \bibinfo
  {author} {\bibfnamefont {S.}~\bibnamefont {Endo}}, \bibinfo {author}
  {\bibfnamefont {W.~J.}\ \bibnamefont {Huggins}}, \bibinfo {author}
  {\bibfnamefont {Y.}~\bibnamefont {Li}}, \bibinfo {author} {\bibfnamefont
  {J.~R.}\ \bibnamefont {McClean}},\ and\ \bibinfo {author} {\bibfnamefont
  {T.~E.}\ \bibnamefont {O'Brien}},\ }\bibfield  {title} {\bibinfo {title}
  {Quantum error mitigation},\ }\href
  {https://doi.org/10.1103/RevModPhys.95.045005} {\bibfield  {journal}
  {\bibinfo  {journal} {Rev. Mod. Phys.}\ }\textbf {\bibinfo {volume} {95}},\
  \bibinfo {pages} {045005} (\bibinfo {year} {2023})}\BibitemShut {NoStop}%
\bibitem [{\citenamefont {Giurgica-Tiron}\ \emph {et~al.}(2020)\citenamefont
  {Giurgica-Tiron}, \citenamefont {Hindy}, \citenamefont {LaRose},
  \citenamefont {Mari},\ and\ \citenamefont {Zeng}}]{giurgica2020digital}%
  \BibitemOpen
  \bibfield  {author} {\bibinfo {author} {\bibfnamefont {T.}~\bibnamefont
  {Giurgica-Tiron}}, \bibinfo {author} {\bibfnamefont {Y.}~\bibnamefont
  {Hindy}}, \bibinfo {author} {\bibfnamefont {R.}~\bibnamefont {LaRose}},
  \bibinfo {author} {\bibfnamefont {A.}~\bibnamefont {Mari}},\ and\ \bibinfo
  {author} {\bibfnamefont {W.~J.}\ \bibnamefont {Zeng}},\ }\bibfield  {title}
  {\bibinfo {title} {Digital zero noise extrapolation for quantum error
  mitigation},\ }in\ \href@noop {} {\emph {\bibinfo {booktitle} {2020 IEEE
  International Conference on Quantum Computing and Engineering (QCE)}}}\
  (\bibinfo {organization} {IEEE},\ \bibinfo {year} {2020})\ pp.\ \bibinfo
  {pages} {306--316}\BibitemShut {NoStop}%
\bibitem [{\citenamefont {Pascuzzi}\ \emph {et~al.}(2022)\citenamefont
  {Pascuzzi}, \citenamefont {He}, \citenamefont {Bauer}, \citenamefont
  {De~Jong},\ and\ \citenamefont {Nachman}}]{pascuzzi2022computationally}%
  \BibitemOpen
  \bibfield  {author} {\bibinfo {author} {\bibfnamefont {V.~R.}\ \bibnamefont
  {Pascuzzi}}, \bibinfo {author} {\bibfnamefont {A.}~\bibnamefont {He}},
  \bibinfo {author} {\bibfnamefont {C.~W.}\ \bibnamefont {Bauer}}, \bibinfo
  {author} {\bibfnamefont {W.~A.}\ \bibnamefont {De~Jong}},\ and\ \bibinfo
  {author} {\bibfnamefont {B.}~\bibnamefont {Nachman}},\ }\bibfield  {title}
  {\bibinfo {title} {Computationally efficient zero-noise extrapolation for
  quantum-gate-error mitigation},\ }\href@noop {} {\bibfield  {journal}
  {\bibinfo  {journal} {Physical Review A}\ }\textbf {\bibinfo {volume}
  {105}},\ \bibinfo {pages} {042406} (\bibinfo {year} {2022})}\BibitemShut
  {NoStop}%
\bibitem [{\citenamefont {He}\ \emph {et~al.}(2020)\citenamefont {He},
  \citenamefont {Nachman}, \citenamefont {de~Jong},\ and\ \citenamefont
  {Bauer}}]{he2020zero}%
  \BibitemOpen
  \bibfield  {author} {\bibinfo {author} {\bibfnamefont {A.}~\bibnamefont
  {He}}, \bibinfo {author} {\bibfnamefont {B.}~\bibnamefont {Nachman}},
  \bibinfo {author} {\bibfnamefont {W.~A.}\ \bibnamefont {de~Jong}},\ and\
  \bibinfo {author} {\bibfnamefont {C.~W.}\ \bibnamefont {Bauer}},\ }\bibfield
  {title} {\bibinfo {title} {Zero-noise extrapolation for quantum-gate error
  mitigation with identity insertions},\ }\href@noop {} {\bibfield  {journal}
  {\bibinfo  {journal} {Physical Review A}\ }\textbf {\bibinfo {volume}
  {102}},\ \bibinfo {pages} {012426} (\bibinfo {year} {2020})}\BibitemShut
  {NoStop}%
\bibitem [{\citenamefont {Mari}\ \emph {et~al.}(2021)\citenamefont {Mari},
  \citenamefont {Shammah},\ and\ \citenamefont {Zeng}}]{mari2021extending}%
  \BibitemOpen
  \bibfield  {author} {\bibinfo {author} {\bibfnamefont {A.}~\bibnamefont
  {Mari}}, \bibinfo {author} {\bibfnamefont {N.}~\bibnamefont {Shammah}},\ and\
  \bibinfo {author} {\bibfnamefont {W.~J.}\ \bibnamefont {Zeng}},\ }\bibfield
  {title} {\bibinfo {title} {Extending quantum probabilistic error cancellation
  by noise scaling},\ }\href@noop {} {\bibfield  {journal} {\bibinfo  {journal}
  {Physical Review A}\ }\textbf {\bibinfo {volume} {104}},\ \bibinfo {pages}
  {052607} (\bibinfo {year} {2021})}\BibitemShut {NoStop}%
\bibitem [{\citenamefont {Van Den~Berg}\ \emph {et~al.}(2023)\citenamefont {Van
  Den~Berg}, \citenamefont {Minev}, \citenamefont {Kandala},\ and\
  \citenamefont {Temme}}]{van2023probabilistic}%
  \BibitemOpen
  \bibfield  {author} {\bibinfo {author} {\bibfnamefont {E.}~\bibnamefont {Van
  Den~Berg}}, \bibinfo {author} {\bibfnamefont {Z.~K.}\ \bibnamefont {Minev}},
  \bibinfo {author} {\bibfnamefont {A.}~\bibnamefont {Kandala}},\ and\ \bibinfo
  {author} {\bibfnamefont {K.}~\bibnamefont {Temme}},\ }\bibfield  {title}
  {\bibinfo {title} {Probabilistic error cancellation with sparse
  pauli--lindblad models on noisy quantum processors},\ }\href@noop {}
  {\bibfield  {journal} {\bibinfo  {journal} {Nature physics}\ }\textbf
  {\bibinfo {volume} {19}},\ \bibinfo {pages} {1116} (\bibinfo {year}
  {2023})}\BibitemShut {NoStop}%
\bibitem [{\citenamefont {Gupta}\ \emph {et~al.}(2024)\citenamefont {Gupta},
  \citenamefont {Van Den~Berg}, \citenamefont {Takita}, \citenamefont {Riste},
  \citenamefont {Temme},\ and\ \citenamefont
  {Kandala}}]{gupta2024probabilistic}%
  \BibitemOpen
  \bibfield  {author} {\bibinfo {author} {\bibfnamefont {R.~S.}\ \bibnamefont
  {Gupta}}, \bibinfo {author} {\bibfnamefont {E.}~\bibnamefont {Van Den~Berg}},
  \bibinfo {author} {\bibfnamefont {M.}~\bibnamefont {Takita}}, \bibinfo
  {author} {\bibfnamefont {D.}~\bibnamefont {Riste}}, \bibinfo {author}
  {\bibfnamefont {K.}~\bibnamefont {Temme}},\ and\ \bibinfo {author}
  {\bibfnamefont {A.}~\bibnamefont {Kandala}},\ }\bibfield  {title} {\bibinfo
  {title} {Probabilistic error cancellation for dynamic quantum circuits},\
  }\href@noop {} {\bibfield  {journal} {\bibinfo  {journal} {Physical Review
  A}\ }\textbf {\bibinfo {volume} {109}},\ \bibinfo {pages} {062617} (\bibinfo
  {year} {2024})}\BibitemShut {NoStop}%
\bibitem [{\citenamefont {Camilo}\ \emph {et~al.}(2025)\citenamefont {Camilo},
  \citenamefont {Maciel}, \citenamefont {Tosta}, \citenamefont {Alhajri},
  \citenamefont {Silva}, \citenamefont {Fran{\c{c}}a},\ and\ \citenamefont
  {Aolita}}]{camilo2025compilation}%
  \BibitemOpen
  \bibfield  {author} {\bibinfo {author} {\bibfnamefont {G.}~\bibnamefont
  {Camilo}}, \bibinfo {author} {\bibfnamefont {T.~O.}\ \bibnamefont {Maciel}},
  \bibinfo {author} {\bibfnamefont {A.}~\bibnamefont {Tosta}}, \bibinfo
  {author} {\bibfnamefont {A.}~\bibnamefont {Alhajri}}, \bibinfo {author}
  {\bibfnamefont {T.~d.~L.}\ \bibnamefont {Silva}}, \bibinfo {author}
  {\bibfnamefont {D.~S.}\ \bibnamefont {Fran{\c{c}}a}},\ and\ \bibinfo {author}
  {\bibfnamefont {L.}~\bibnamefont {Aolita}},\ }\bibfield  {title} {\bibinfo
  {title} {Compilation-informed probabilistic quantum error cancellation},\
  }\href@noop {} {\bibfield  {journal} {\bibinfo  {journal} {arXiv preprint
  arXiv:2508.20174}\ } (\bibinfo {year} {2025})}\BibitemShut {NoStop}%
\bibitem [{\citenamefont {Mooney}\ \emph {et~al.}(2021)\citenamefont {Mooney},
  \citenamefont {White}, \citenamefont {Hill},\ and\ \citenamefont
  {Hollenberg}}]{mooney2021whole}%
  \BibitemOpen
  \bibfield  {author} {\bibinfo {author} {\bibfnamefont {G.~J.}\ \bibnamefont
  {Mooney}}, \bibinfo {author} {\bibfnamefont {G.~A.}\ \bibnamefont {White}},
  \bibinfo {author} {\bibfnamefont {C.~D.}\ \bibnamefont {Hill}},\ and\
  \bibinfo {author} {\bibfnamefont {L.~C.}\ \bibnamefont {Hollenberg}},\
  }\bibfield  {title} {\bibinfo {title} {Whole-device entanglement in a
  65-qubit superconducting quantum computer},\ }\href@noop {} {\bibfield
  {journal} {\bibinfo  {journal} {Advanced Quantum Technologies}\ }\textbf
  {\bibinfo {volume} {4}},\ \bibinfo {pages} {2100061} (\bibinfo {year}
  {2021})}\BibitemShut {NoStop}%
\bibitem [{\citenamefont {Nation}\ \emph {et~al.}(2021)\citenamefont {Nation},
  \citenamefont {Kang}, \citenamefont {Sundaresan},\ and\ \citenamefont
  {Gambetta}}]{nation2021scalable}%
  \BibitemOpen
  \bibfield  {author} {\bibinfo {author} {\bibfnamefont {P.~D.}\ \bibnamefont
  {Nation}}, \bibinfo {author} {\bibfnamefont {H.}~\bibnamefont {Kang}},
  \bibinfo {author} {\bibfnamefont {N.}~\bibnamefont {Sundaresan}},\ and\
  \bibinfo {author} {\bibfnamefont {J.~M.}\ \bibnamefont {Gambetta}},\
  }\bibfield  {title} {\bibinfo {title} {Scalable mitigation of measurement
  errors on quantum computers},\ }\href@noop {} {\bibfield  {journal} {\bibinfo
   {journal} {PRX Quantum}\ }\textbf {\bibinfo {volume} {2}},\ \bibinfo {pages}
  {040326} (\bibinfo {year} {2021})}\BibitemShut {NoStop}%
\bibitem [{\citenamefont {Zhu}\ \emph {et~al.}(2019)\citenamefont {Zhu},
  \citenamefont {Linke}, \citenamefont {Benedetti}, \citenamefont {Landsman},
  \citenamefont {Nguyen}, \citenamefont {Alderete}, \citenamefont
  {Perdomo-Ortiz}, \citenamefont {Korda}, \citenamefont {Garfoot},
  \citenamefont {Brecque} \emph {et~al.}}]{zhu2019training}%
  \BibitemOpen
  \bibfield  {author} {\bibinfo {author} {\bibfnamefont {D.}~\bibnamefont
  {Zhu}}, \bibinfo {author} {\bibfnamefont {N.~M.}\ \bibnamefont {Linke}},
  \bibinfo {author} {\bibfnamefont {M.}~\bibnamefont {Benedetti}}, \bibinfo
  {author} {\bibfnamefont {K.~A.}\ \bibnamefont {Landsman}}, \bibinfo {author}
  {\bibfnamefont {N.~H.}\ \bibnamefont {Nguyen}}, \bibinfo {author}
  {\bibfnamefont {C.~H.}\ \bibnamefont {Alderete}}, \bibinfo {author}
  {\bibfnamefont {A.}~\bibnamefont {Perdomo-Ortiz}}, \bibinfo {author}
  {\bibfnamefont {N.}~\bibnamefont {Korda}}, \bibinfo {author} {\bibfnamefont
  {A.}~\bibnamefont {Garfoot}}, \bibinfo {author} {\bibfnamefont
  {C.}~\bibnamefont {Brecque}}, \emph {et~al.},\ }\bibfield  {title} {\bibinfo
  {title} {Training of quantum circuits on a hybrid quantum computer},\
  }\href@noop {} {\bibfield  {journal} {\bibinfo  {journal} {Science advances}\
  }\textbf {\bibinfo {volume} {5}},\ \bibinfo {pages} {eaaw9918} (\bibinfo
  {year} {2019})}\BibitemShut {NoStop}%
\bibitem [{\citenamefont {Quantum}\ \emph {et~al.}(2020)\citenamefont
  {Quantum}, \citenamefont {Collaborators*†}, \citenamefont {Arute},
  \citenamefont {Arya}, \citenamefont {Babbush}, \citenamefont {Bacon},
  \citenamefont {Bardin}, \citenamefont {Barends}, \citenamefont {Boixo},
  \citenamefont {Broughton}, \citenamefont {Buckley} \emph
  {et~al.}}]{google2020hartree}%
  \BibitemOpen
  \bibfield  {author} {\bibinfo {author} {\bibfnamefont {G.~A.}\ \bibnamefont
  {Quantum}}, \bibinfo {author} {\bibnamefont {Collaborators*†}}, \bibinfo
  {author} {\bibfnamefont {F.}~\bibnamefont {Arute}}, \bibinfo {author}
  {\bibfnamefont {K.}~\bibnamefont {Arya}}, \bibinfo {author} {\bibfnamefont
  {R.}~\bibnamefont {Babbush}}, \bibinfo {author} {\bibfnamefont
  {D.}~\bibnamefont {Bacon}}, \bibinfo {author} {\bibfnamefont {J.~C.}\
  \bibnamefont {Bardin}}, \bibinfo {author} {\bibfnamefont {R.}~\bibnamefont
  {Barends}}, \bibinfo {author} {\bibfnamefont {S.}~\bibnamefont {Boixo}},
  \bibinfo {author} {\bibfnamefont {M.}~\bibnamefont {Broughton}}, \bibinfo
  {author} {\bibfnamefont {B.~B.}\ \bibnamefont {Buckley}}, \emph {et~al.},\
  }\bibfield  {title} {\bibinfo {title} {Hartree-fock on a superconducting
  qubit quantum computer},\ }\href@noop {} {\bibfield  {journal} {\bibinfo
  {journal} {Science}\ }\textbf {\bibinfo {volume} {369}},\ \bibinfo {pages}
  {1084} (\bibinfo {year} {2020})}\BibitemShut {NoStop}%
\bibitem [{\citenamefont {Cerezo}\ \emph {et~al.}(2021)\citenamefont {Cerezo},
  \citenamefont {Arrasmith}, \citenamefont {Babbush}, \citenamefont {Benjamin},
  \citenamefont {Endo}, \citenamefont {Fujii}, \citenamefont {McClean},
  \citenamefont {Mitarai}, \citenamefont {Yuan}, \citenamefont {Cincio} \emph
  {et~al.}}]{cerezo2021variational}%
  \BibitemOpen
  \bibfield  {author} {\bibinfo {author} {\bibfnamefont {M.}~\bibnamefont
  {Cerezo}}, \bibinfo {author} {\bibfnamefont {A.}~\bibnamefont {Arrasmith}},
  \bibinfo {author} {\bibfnamefont {R.}~\bibnamefont {Babbush}}, \bibinfo
  {author} {\bibfnamefont {S.~C.}\ \bibnamefont {Benjamin}}, \bibinfo {author}
  {\bibfnamefont {S.}~\bibnamefont {Endo}}, \bibinfo {author} {\bibfnamefont
  {K.}~\bibnamefont {Fujii}}, \bibinfo {author} {\bibfnamefont {J.~R.}\
  \bibnamefont {McClean}}, \bibinfo {author} {\bibfnamefont {K.}~\bibnamefont
  {Mitarai}}, \bibinfo {author} {\bibfnamefont {X.}~\bibnamefont {Yuan}},
  \bibinfo {author} {\bibfnamefont {L.}~\bibnamefont {Cincio}}, \emph
  {et~al.},\ }\bibfield  {title} {\bibinfo {title} {Variational quantum
  algorithms},\ }\href@noop {} {\bibfield  {journal} {\bibinfo  {journal}
  {Nature Reviews Physics}\ }\textbf {\bibinfo {volume} {3}},\ \bibinfo {pages}
  {625} (\bibinfo {year} {2021})}\BibitemShut {NoStop}%
\bibitem [{\citenamefont {Ostaszewski}\ \emph {et~al.}(2021)\citenamefont
  {Ostaszewski}, \citenamefont {Grant},\ and\ \citenamefont
  {Benedetti}}]{ostaszewski2021structure}%
  \BibitemOpen
  \bibfield  {author} {\bibinfo {author} {\bibfnamefont {M.}~\bibnamefont
  {Ostaszewski}}, \bibinfo {author} {\bibfnamefont {E.}~\bibnamefont {Grant}},\
  and\ \bibinfo {author} {\bibfnamefont {M.}~\bibnamefont {Benedetti}},\
  }\bibfield  {title} {\bibinfo {title} {Structure optimization for
  parameterized quantum circuits},\ }\href@noop {} {\bibfield  {journal}
  {\bibinfo  {journal} {Quantum}\ }\textbf {\bibinfo {volume} {5}},\ \bibinfo
  {pages} {391} (\bibinfo {year} {2021})}\BibitemShut {NoStop}%
\bibitem [{\citenamefont {Chen}\ \emph
  {et~al.}(2023{\natexlab{b}})\citenamefont {Chen}, \citenamefont {Shen},
  \citenamefont {Lee},\ and\ \citenamefont {Yang}}]{chen2022high}%
  \BibitemOpen
  \bibfield  {author} {\bibinfo {author} {\bibfnamefont {T.}~\bibnamefont
  {Chen}}, \bibinfo {author} {\bibfnamefont {R.}~\bibnamefont {Shen}}, \bibinfo
  {author} {\bibfnamefont {C.~H.}\ \bibnamefont {Lee}},\ and\ \bibinfo {author}
  {\bibfnamefont {B.}~\bibnamefont {Yang}},\ }\bibfield  {title} {\bibinfo
  {title} {High-fidelity realization of the aklt state on a nisq-era quantum
  processor},\ }\href@noop {} {\bibfield  {journal} {\bibinfo  {journal}
  {SciPost Physics}\ }\textbf {\bibinfo {volume} {15}},\ \bibinfo {pages} {170}
  (\bibinfo {year} {2023}{\natexlab{b}})}\BibitemShut {NoStop}%
\bibitem [{\citenamefont {Koh}\ \emph {et~al.}(2023)\citenamefont {Koh},
  \citenamefont {Tai},\ and\ \citenamefont {Lee}}]{koh2023observation}%
  \BibitemOpen
  \bibfield  {author} {\bibinfo {author} {\bibfnamefont {J.~M.}\ \bibnamefont
  {Koh}}, \bibinfo {author} {\bibfnamefont {T.}~\bibnamefont {Tai}},\ and\
  \bibinfo {author} {\bibfnamefont {C.~H.}\ \bibnamefont {Lee}},\ }\bibfield
  {title} {\bibinfo {title} {Observation of higher-order topological states on
  a quantum computer},\ }\href@noop {} {\bibfield  {journal} {\bibinfo
  {journal} {arXiv preprint arXiv:2303.02179}\ } (\bibinfo {year}
  {2023})}\BibitemShut {NoStop}%
\bibitem [{\citenamefont {Shen}\ and\ \citenamefont {Lee}(2022)}]{shen2022non}%
  \BibitemOpen
  \bibfield  {author} {\bibinfo {author} {\bibfnamefont {R.}~\bibnamefont
  {Shen}}\ and\ \bibinfo {author} {\bibfnamefont {C.~H.}\ \bibnamefont {Lee}},\
  }\bibfield  {title} {\bibinfo {title} {Non-hermitian skin clusters from
  strong interactions},\ }\href@noop {} {\bibfield  {journal} {\bibinfo
  {journal} {Communications Physics}\ }\textbf {\bibinfo {volume} {5}},\
  \bibinfo {pages} {1} (\bibinfo {year} {2022})}\BibitemShut {NoStop}%
\bibitem [{\citenamefont {Qin}\ \emph {et~al.}(2022)\citenamefont {Qin},
  \citenamefont {Ma}, \citenamefont {Shen},\ and\ \citenamefont
  {Lee}}]{qin2022universal}%
  \BibitemOpen
  \bibfield  {author} {\bibinfo {author} {\bibfnamefont {F.}~\bibnamefont
  {Qin}}, \bibinfo {author} {\bibfnamefont {Y.}~\bibnamefont {Ma}}, \bibinfo
  {author} {\bibfnamefont {R.}~\bibnamefont {Shen}},\ and\ \bibinfo {author}
  {\bibfnamefont {C.~H.}\ \bibnamefont {Lee}},\ }\bibfield  {title} {\bibinfo
  {title} {Universal competitive spectral scaling from the critical
  non-hermitian skin effect},\ }\href@noop {} {\bibfield  {journal} {\bibinfo
  {journal} {arXiv preprint arXiv:2212.13536}\ } (\bibinfo {year}
  {2022})}\BibitemShut {NoStop}%
\bibitem [{\citenamefont {Liu}\ \emph {et~al.}(2021)\citenamefont {Liu},
  \citenamefont {Zhang},\ and\ \citenamefont {Chen}}]{liu2021non}%
  \BibitemOpen
  \bibfield  {author} {\bibinfo {author} {\bibfnamefont {C.}~\bibnamefont
  {Liu}}, \bibinfo {author} {\bibfnamefont {P.}~\bibnamefont {Zhang}},\ and\
  \bibinfo {author} {\bibfnamefont {X.}~\bibnamefont {Chen}},\ }\bibfield
  {title} {\bibinfo {title} {Non-unitary dynamics of sachdev-ye-kitaev chain},\
  }\href@noop {} {\bibfield  {journal} {\bibinfo  {journal} {SciPost Physics}\
  }\textbf {\bibinfo {volume} {10}},\ \bibinfo {pages} {048} (\bibinfo {year}
  {2021})}\BibitemShut {NoStop}%
\bibitem [{\citenamefont {Yang}\ \emph {et~al.}(2024)\citenamefont {Yang},
  \citenamefont {Yuan},\ and\ \citenamefont {Lee}}]{yang2024non}%
  \BibitemOpen
  \bibfield  {author} {\bibinfo {author} {\bibfnamefont {M.}~\bibnamefont
  {Yang}}, \bibinfo {author} {\bibfnamefont {L.}~\bibnamefont {Yuan}},\ and\
  \bibinfo {author} {\bibfnamefont {C.~H.}\ \bibnamefont {Lee}},\ }\bibfield
  {title} {\bibinfo {title} {Non-hermitian ultra-strong bosonic condensation
  through interaction-induced caging},\ }\href@noop {} {\bibfield  {journal}
  {\bibinfo  {journal} {arXiv preprint arXiv:2410.01258}\ } (\bibinfo {year}
  {2024})}\BibitemShut {NoStop}%
\bibitem [{\citenamefont {Qin}\ \emph {et~al.}(2024)\citenamefont {Qin},
  \citenamefont {Shen}, \citenamefont {Li},\ and\ \citenamefont
  {Lee}}]{qin2024kinked}%
  \BibitemOpen
  \bibfield  {author} {\bibinfo {author} {\bibfnamefont {F.}~\bibnamefont
  {Qin}}, \bibinfo {author} {\bibfnamefont {R.}~\bibnamefont {Shen}}, \bibinfo
  {author} {\bibfnamefont {L.}~\bibnamefont {Li}},\ and\ \bibinfo {author}
  {\bibfnamefont {C.~H.}\ \bibnamefont {Lee}},\ }\bibfield  {title} {\bibinfo
  {title} {Kinked linear response from non-hermitian cold-atom pumping},\
  }\href@noop {} {\bibfield  {journal} {\bibinfo  {journal} {Physical Review
  A}\ }\textbf {\bibinfo {volume} {109}},\ \bibinfo {pages} {053311} (\bibinfo
  {year} {2024})}\BibitemShut {NoStop}%
\bibitem [{\citenamefont {Zhang}\ \emph {et~al.}(2022)\citenamefont {Zhang},
  \citenamefont {Liu}, \citenamefont {Jian},\ and\ \citenamefont
  {Chen}}]{zhang2022universal}%
  \BibitemOpen
  \bibfield  {author} {\bibinfo {author} {\bibfnamefont {P.}~\bibnamefont
  {Zhang}}, \bibinfo {author} {\bibfnamefont {C.}~\bibnamefont {Liu}}, \bibinfo
  {author} {\bibfnamefont {S.-K.}\ \bibnamefont {Jian}},\ and\ \bibinfo
  {author} {\bibfnamefont {X.}~\bibnamefont {Chen}},\ }\bibfield  {title}
  {\bibinfo {title} {Universal entanglement transitions of free fermions with
  long-range non-unitary dynamics},\ }\href@noop {} {\bibfield  {journal}
  {\bibinfo  {journal} {Quantum}\ }\textbf {\bibinfo {volume} {6}},\ \bibinfo
  {pages} {723} (\bibinfo {year} {2022})}\BibitemShut {NoStop}%
\bibitem [{\citenamefont {Kawabata}\ \emph {et~al.}(2022)\citenamefont
  {Kawabata}, \citenamefont {Shiozaki},\ and\ \citenamefont
  {Ryu}}]{kawabata2022many}%
  \BibitemOpen
  \bibfield  {author} {\bibinfo {author} {\bibfnamefont {K.}~\bibnamefont
  {Kawabata}}, \bibinfo {author} {\bibfnamefont {K.}~\bibnamefont {Shiozaki}},\
  and\ \bibinfo {author} {\bibfnamefont {S.}~\bibnamefont {Ryu}},\ }\bibfield
  {title} {\bibinfo {title} {Many-body topology of non-hermitian systems},\
  }\href@noop {} {\bibfield  {journal} {\bibinfo  {journal} {Physical Review
  B}\ }\textbf {\bibinfo {volume} {105}},\ \bibinfo {pages} {165137} (\bibinfo
  {year} {2022})}\BibitemShut {NoStop}%
\bibitem [{\citenamefont {Jiang}\ and\ \citenamefont
  {Lee}(2022)}]{jiang2022dimensional}%
  \BibitemOpen
  \bibfield  {author} {\bibinfo {author} {\bibfnamefont {H.}~\bibnamefont
  {Jiang}}\ and\ \bibinfo {author} {\bibfnamefont {C.~H.}\ \bibnamefont
  {Lee}},\ }\bibfield  {title} {\bibinfo {title} {Dimensional transmutation
  from non-hermiticity},\ }\href@noop {} {\bibfield  {journal} {\bibinfo
  {journal} {arXiv preprint arXiv:2207.08843}\ } (\bibinfo {year}
  {2022})}\BibitemShut {NoStop}%
\bibitem [{\citenamefont {Yoshida}\ and\ \citenamefont
  {Hatsugai}(2023)}]{yoshida2023fate}%
  \BibitemOpen
  \bibfield  {author} {\bibinfo {author} {\bibfnamefont {T.}~\bibnamefont
  {Yoshida}}\ and\ \bibinfo {author} {\bibfnamefont {Y.}~\bibnamefont
  {Hatsugai}},\ }\bibfield  {title} {\bibinfo {title} {Fate of exceptional
  points under interactions: Reduction of topological classifications},\
  }\href@noop {} {\bibfield  {journal} {\bibinfo  {journal} {Physical Review
  B}\ }\textbf {\bibinfo {volume} {107}},\ \bibinfo {pages} {075118} (\bibinfo
  {year} {2023})}\BibitemShut {NoStop}%
\bibitem [{\citenamefont {Poddubny}(2023)}]{poddubny2023interaction}%
  \BibitemOpen
  \bibfield  {author} {\bibinfo {author} {\bibfnamefont {A.~N.}\ \bibnamefont
  {Poddubny}},\ }\bibfield  {title} {\bibinfo {title} {Interaction-induced
  analog of a non-hermitian skin effect in a lattice two-body problem},\
  }\href@noop {} {\bibfield  {journal} {\bibinfo  {journal} {Physical Review
  B}\ }\textbf {\bibinfo {volume} {107}},\ \bibinfo {pages} {045131} (\bibinfo
  {year} {2023})}\BibitemShut {NoStop}%
\bibitem [{\citenamefont {Shen}\ \emph
  {et~al.}(2025{\natexlab{b}})\citenamefont {Shen}, \citenamefont {Chan},\ and\
  \citenamefont {Lee}}]{shen2025non}%
  \BibitemOpen
  \bibfield  {author} {\bibinfo {author} {\bibfnamefont {R.}~\bibnamefont
  {Shen}}, \bibinfo {author} {\bibfnamefont {W.~J.}\ \bibnamefont {Chan}},\
  and\ \bibinfo {author} {\bibfnamefont {C.~H.}\ \bibnamefont {Lee}},\
  }\bibfield  {title} {\bibinfo {title} {Non-hermitian skin effect along
  hyperbolic geodesics},\ }\href@noop {} {\bibfield  {journal} {\bibinfo
  {journal} {Physical Review B}\ }\textbf {\bibinfo {volume} {111}},\ \bibinfo
  {pages} {045420} (\bibinfo {year} {2025}{\natexlab{b}})}\BibitemShut
  {NoStop}%
\bibitem [{\citenamefont {Fu}\ and\ \citenamefont
  {Zhang}(2023)}]{fu2023anatomy}%
  \BibitemOpen
  \bibfield  {author} {\bibinfo {author} {\bibfnamefont {Y.}~\bibnamefont
  {Fu}}\ and\ \bibinfo {author} {\bibfnamefont {Y.}~\bibnamefont {Zhang}},\
  }\bibfield  {title} {\bibinfo {title} {Anatomy of open-boundary bulk in
  multiband non-hermitian systems},\ }\href@noop {} {\bibfield  {journal}
  {\bibinfo  {journal} {Physical Review B}\ }\textbf {\bibinfo {volume}
  {107}},\ \bibinfo {pages} {115412} (\bibinfo {year} {2023})}\BibitemShut
  {NoStop}%
\bibitem [{\citenamefont {Li}\ \emph {et~al.}(2025)\citenamefont {Li},
  \citenamefont {Jiang},\ and\ \citenamefont {Lee}}]{li2025phase}%
  \BibitemOpen
  \bibfield  {author} {\bibinfo {author} {\bibfnamefont {Q.}~\bibnamefont
  {Li}}, \bibinfo {author} {\bibfnamefont {H.}~\bibnamefont {Jiang}},\ and\
  \bibinfo {author} {\bibfnamefont {C.~H.}\ \bibnamefont {Lee}},\ }\bibfield
  {title} {\bibinfo {title} {Phase-space generalized brillouin zone for
  spatially inhomogeneous non-hermitian systems},\ }\href@noop {} {\bibfield
  {journal} {\bibinfo  {journal} {arXiv preprint arXiv:2501.09785}\ } (\bibinfo
  {year} {2025})}\BibitemShut {NoStop}%
\bibitem [{\citenamefont {Shen}\ \emph {et~al.}(2023)\citenamefont {Shen},
  \citenamefont {Chen}, \citenamefont {Qin}, \citenamefont {Zhong},\ and\
  \citenamefont {Lee}}]{shen2023proposal}%
  \BibitemOpen
  \bibfield  {author} {\bibinfo {author} {\bibfnamefont {R.}~\bibnamefont
  {Shen}}, \bibinfo {author} {\bibfnamefont {T.}~\bibnamefont {Chen}}, \bibinfo
  {author} {\bibfnamefont {F.}~\bibnamefont {Qin}}, \bibinfo {author}
  {\bibfnamefont {Y.}~\bibnamefont {Zhong}},\ and\ \bibinfo {author}
  {\bibfnamefont {C.~H.}\ \bibnamefont {Lee}},\ }\bibfield  {title} {\bibinfo
  {title} {Proposal for observing yang-lee criticality in rydberg atomic
  arrays},\ }\href@noop {} {\bibfield  {journal} {\bibinfo  {journal} {arXiv
  preprint arXiv:2302.06662}\ } (\bibinfo {year} {2023})}\BibitemShut {NoStop}%
\bibitem [{\citenamefont {Lee}\ \emph {et~al.}(2015)\citenamefont {Lee},
  \citenamefont {Papi\ifmmode~\acute{c}\else \'{c}\fi{}},\ and\ \citenamefont
  {Thomale}}]{PhysRevX.5.041003}%
  \BibitemOpen
  \bibfield  {author} {\bibinfo {author} {\bibfnamefont {C.~H.}\ \bibnamefont
  {Lee}}, \bibinfo {author} {\bibfnamefont {Z.}~\bibnamefont
  {Papi\ifmmode~\acute{c}\else \'{c}\fi{}}},\ and\ \bibinfo {author}
  {\bibfnamefont {R.}~\bibnamefont {Thomale}},\ }\bibfield  {title} {\bibinfo
  {title} {Geometric construction of quantum hall clustering hamiltonians},\
  }\href {https://doi.org/10.1103/PhysRevX.5.041003} {\bibfield  {journal}
  {\bibinfo  {journal} {Phys. Rev. X}\ }\textbf {\bibinfo {volume} {5}},\
  \bibinfo {pages} {041003} (\bibinfo {year} {2015})}\BibitemShut {NoStop}%
\bibitem [{\citenamefont {Lee}\ \emph {et~al.}(2018)\citenamefont {Lee},
  \citenamefont {Ho}, \citenamefont {Yang}, \citenamefont {Gong},\ and\
  \citenamefont {Papi\ifmmode~\acute{c}\else
  \'{c}\fi{}}}]{PhysRevLett.121.237401}%
  \BibitemOpen
  \bibfield  {author} {\bibinfo {author} {\bibfnamefont {C.~H.}\ \bibnamefont
  {Lee}}, \bibinfo {author} {\bibfnamefont {W.~W.}\ \bibnamefont {Ho}},
  \bibinfo {author} {\bibfnamefont {B.}~\bibnamefont {Yang}}, \bibinfo {author}
  {\bibfnamefont {J.}~\bibnamefont {Gong}},\ and\ \bibinfo {author}
  {\bibfnamefont {Z.}~\bibnamefont {Papi\ifmmode~\acute{c}\else \'{c}\fi{}}},\
  }\bibfield  {title} {\bibinfo {title} {Floquet mechanism for non-abelian
  fractional quantum hall states},\ }\href
  {https://doi.org/10.1103/PhysRevLett.121.237401} {\bibfield  {journal}
  {\bibinfo  {journal} {Phys. Rev. Lett.}\ }\textbf {\bibinfo {volume} {121}},\
  \bibinfo {pages} {237401} (\bibinfo {year} {2018})}\BibitemShut {NoStop}%
\bibitem [{\citenamefont {Qin}\ \emph {et~al.}(2025)\citenamefont {Qin},
  \citenamefont {Lee},\ and\ \citenamefont {Li}}]{qin2025dynamical}%
  \BibitemOpen
  \bibfield  {author} {\bibinfo {author} {\bibfnamefont {Y.}~\bibnamefont
  {Qin}}, \bibinfo {author} {\bibfnamefont {C.~H.}\ \bibnamefont {Lee}},\ and\
  \bibinfo {author} {\bibfnamefont {L.}~\bibnamefont {Li}},\ }\bibfield
  {title} {\bibinfo {title} {Dynamical suppression of many-body non-hermitian
  skin effect in anyonic systems},\ }\href@noop {} {\bibfield  {journal}
  {\bibinfo  {journal} {Communications Physics}\ }\textbf {\bibinfo {volume}
  {8}},\ \bibinfo {pages} {18} (\bibinfo {year} {2025})}\BibitemShut {NoStop}%
\bibitem [{\citenamefont {Lapierre}\ \emph {et~al.}(2025)\citenamefont
  {Lapierre}, \citenamefont {Pelliconi}, \citenamefont {Ryu},\ and\
  \citenamefont {Sonner}}]{lapierre2025driven}%
  \BibitemOpen
  \bibfield  {author} {\bibinfo {author} {\bibfnamefont {B.}~\bibnamefont
  {Lapierre}}, \bibinfo {author} {\bibfnamefont {P.}~\bibnamefont {Pelliconi}},
  \bibinfo {author} {\bibfnamefont {S.}~\bibnamefont {Ryu}},\ and\ \bibinfo
  {author} {\bibfnamefont {J.}~\bibnamefont {Sonner}},\ }\bibfield  {title}
  {\bibinfo {title} {Driven non-unitary dynamics of quantum critical systems},\
  }\href@noop {} {\bibfield  {journal} {\bibinfo  {journal} {arXiv preprint
  arXiv:2505.01508}\ } (\bibinfo {year} {2025})}\BibitemShut {NoStop}%
\bibitem [{\citenamefont {Schuld}\ \emph {et~al.}(2014)\citenamefont {Schuld},
  \citenamefont {Sinayskiy},\ and\ \citenamefont
  {Petruccione}}]{schuld2014quest}%
  \BibitemOpen
  \bibfield  {author} {\bibinfo {author} {\bibfnamefont {M.}~\bibnamefont
  {Schuld}}, \bibinfo {author} {\bibfnamefont {I.}~\bibnamefont {Sinayskiy}},\
  and\ \bibinfo {author} {\bibfnamefont {F.}~\bibnamefont {Petruccione}},\
  }\bibfield  {title} {\bibinfo {title} {The quest for a quantum neural
  network},\ }\href@noop {} {\bibfield  {journal} {\bibinfo  {journal} {Quantum
  Information Processing}\ }\textbf {\bibinfo {volume} {13}},\ \bibinfo {pages}
  {2567} (\bibinfo {year} {2014})}\BibitemShut {NoStop}%
\bibitem [{\citenamefont {Jeswal}\ and\ \citenamefont
  {Chakraverty}(2019)}]{jeswal2019recent}%
  \BibitemOpen
  \bibfield  {author} {\bibinfo {author} {\bibfnamefont {S.}~\bibnamefont
  {Jeswal}}\ and\ \bibinfo {author} {\bibfnamefont {S.}~\bibnamefont
  {Chakraverty}},\ }\bibfield  {title} {\bibinfo {title} {Recent developments
  and applications in quantum neural network: A review},\ }\href@noop {}
  {\bibfield  {journal} {\bibinfo  {journal} {Archives of Computational Methods
  in Engineering}\ }\textbf {\bibinfo {volume} {26}},\ \bibinfo {pages} {793}
  (\bibinfo {year} {2019})}\BibitemShut {NoStop}%
\bibitem [{\citenamefont {Altaisky}(2001)}]{altaisky2001quantum}%
  \BibitemOpen
  \bibfield  {author} {\bibinfo {author} {\bibfnamefont {M.}~\bibnamefont
  {Altaisky}},\ }\bibfield  {title} {\bibinfo {title} {Quantum neural
  network},\ }\href@noop {} {\bibfield  {journal} {\bibinfo  {journal} {arXiv
  preprint quant-ph/0107012}\ } (\bibinfo {year} {2001})}\BibitemShut {NoStop}%
\bibitem [{\citenamefont {Ricks}\ and\ \citenamefont
  {Ventura}(2003)}]{ricks2003training}%
  \BibitemOpen
  \bibfield  {author} {\bibinfo {author} {\bibfnamefont {B.}~\bibnamefont
  {Ricks}}\ and\ \bibinfo {author} {\bibfnamefont {D.}~\bibnamefont
  {Ventura}},\ }\bibfield  {title} {\bibinfo {title} {Training a quantum neural
  network},\ }\href@noop {} {\bibfield  {journal} {\bibinfo  {journal}
  {Advances in neural information processing systems}\ }\textbf {\bibinfo
  {volume} {16}} (\bibinfo {year} {2003})}\BibitemShut {NoStop}%
\bibitem [{\citenamefont {Rodriguez-Vega}\ \emph {et~al.}(2022)\citenamefont
  {Rodriguez-Vega}, \citenamefont {Carlander}, \citenamefont {Bahri},
  \citenamefont {Lin}, \citenamefont {Sinitsyn},\ and\ \citenamefont
  {Fiete}}]{rodriguez2022real}%
  \BibitemOpen
  \bibfield  {author} {\bibinfo {author} {\bibfnamefont {M.}~\bibnamefont
  {Rodriguez-Vega}}, \bibinfo {author} {\bibfnamefont {E.}~\bibnamefont
  {Carlander}}, \bibinfo {author} {\bibfnamefont {A.}~\bibnamefont {Bahri}},
  \bibinfo {author} {\bibfnamefont {Z.-X.}\ \bibnamefont {Lin}}, \bibinfo
  {author} {\bibfnamefont {N.~A.}\ \bibnamefont {Sinitsyn}},\ and\ \bibinfo
  {author} {\bibfnamefont {G.~A.}\ \bibnamefont {Fiete}},\ }\bibfield  {title}
  {\bibinfo {title} {Real-time simulation of light-driven spin chains on
  quantum computers},\ }\href@noop {} {\bibfield  {journal} {\bibinfo
  {journal} {Physical Review Research}\ }\textbf {\bibinfo {volume} {4}},\
  \bibinfo {pages} {013196} (\bibinfo {year} {2022})}\BibitemShut {NoStop}%
\bibitem [{\citenamefont {Robertson}\ and\ \citenamefont
  {Song}(2023)}]{robertson2022mitigating}%
  \BibitemOpen
  \bibfield  {author} {\bibinfo {author} {\bibfnamefont {A.}~\bibnamefont
  {Robertson}}\ and\ \bibinfo {author} {\bibfnamefont {S.}~\bibnamefont
  {Song}},\ }\bibfield  {title} {\bibinfo {title} {Mitigating coupling map
  constrained correlated measurement errors on quantum devices},\ }\bibfield
  {booktitle} {\emph {\bibinfo {booktitle} {Proceedings of the International
  Conference for High Performance Computing, Networking, Storage and
  Analysis}},\ }\href@noop {} {\ ,\ \bibinfo {pages} {1} (\bibinfo {year}
  {2023})}\BibitemShut {NoStop}%
\bibitem [{\citenamefont {{Qiskit contributors}}(2023)}]{Qiskit}%
  \BibitemOpen
  \bibfield  {author} {\bibinfo {author} {\bibnamefont {{Qiskit
  contributors}}},\ }\href {https://doi.org/10.5281/zenodo.2573505} {\bibinfo
  {title} {Qiskit: An open-source framework for quantum computing}} (\bibinfo
  {year} {2023})\BibitemShut {NoStop}%
\bibitem [{\citenamefont {Yang}\ \emph {et~al.}(2017)\citenamefont {Yang},
  \citenamefont {Rahmani}, \citenamefont {Shabani}, \citenamefont {Neven},\
  and\ \citenamefont {Chamon}}]{yang2017optimizing}%
  \BibitemOpen
  \bibfield  {author} {\bibinfo {author} {\bibfnamefont {Z.-C.}\ \bibnamefont
  {Yang}}, \bibinfo {author} {\bibfnamefont {A.}~\bibnamefont {Rahmani}},
  \bibinfo {author} {\bibfnamefont {A.}~\bibnamefont {Shabani}}, \bibinfo
  {author} {\bibfnamefont {H.}~\bibnamefont {Neven}},\ and\ \bibinfo {author}
  {\bibfnamefont {C.}~\bibnamefont {Chamon}},\ }\bibfield  {title} {\bibinfo
  {title} {Optimizing variational quantum algorithms using pontryagin’s
  minimum principle},\ }\href@noop {} {\bibfield  {journal} {\bibinfo
  {journal} {Physical Review X}\ }\textbf {\bibinfo {volume} {7}},\ \bibinfo
  {pages} {021027} (\bibinfo {year} {2017})}\BibitemShut {NoStop}%
\bibitem [{\citenamefont {van Straaten}\ and\ \citenamefont
  {Koczor}(2021)}]{van2021measurement}%
  \BibitemOpen
  \bibfield  {author} {\bibinfo {author} {\bibfnamefont {B.}~\bibnamefont {van
  Straaten}}\ and\ \bibinfo {author} {\bibfnamefont {B.}~\bibnamefont
  {Koczor}},\ }\bibfield  {title} {\bibinfo {title} {Measurement cost of
  metric-aware variational quantum algorithms},\ }\href@noop {} {\bibfield
  {journal} {\bibinfo  {journal} {PRX Quantum}\ }\textbf {\bibinfo {volume}
  {2}},\ \bibinfo {pages} {030324} (\bibinfo {year} {2021})}\BibitemShut
  {NoStop}%
\bibitem [{\citenamefont {Zhou}\ \emph {et~al.}(2025)\citenamefont {Zhou},
  \citenamefont {He}, \citenamefont {Pang}, \citenamefont {Lyu}, \citenamefont
  {Zhang},\ and\ \citenamefont {Chen}}]{PhysRevA.111.042616}%
  \BibitemOpen
  \bibfield  {author} {\bibinfo {author} {\bibfnamefont {Y.}~\bibnamefont
  {Zhou}}, \bibinfo {author} {\bibfnamefont {H.}~\bibnamefont {He}}, \bibinfo
  {author} {\bibfnamefont {F.}~\bibnamefont {Pang}}, \bibinfo {author}
  {\bibfnamefont {H.}~\bibnamefont {Lyu}}, \bibinfo {author} {\bibfnamefont
  {Y.}~\bibnamefont {Zhang}},\ and\ \bibinfo {author} {\bibfnamefont
  {X.}~\bibnamefont {Chen}},\ }\bibfield  {title} {\bibinfo {title}
  {Variational quantum compiling for three-qubit-gate design in quantum dots},\
  }\href {https://doi.org/10.1103/PhysRevA.111.042616} {\bibfield  {journal}
  {\bibinfo  {journal} {Phys. Rev. A}\ }\textbf {\bibinfo {volume} {111}},\
  \bibinfo {pages} {042616} (\bibinfo {year} {2025})}\BibitemShut {NoStop}%
\bibitem [{\citenamefont {Ji}\ \emph {et~al.}(2025)\citenamefont {Ji},
  \citenamefont {Chen}, \citenamefont {Polian},\ and\ \citenamefont
  {Ban}}]{ji2025algorithm}%
  \BibitemOpen
  \bibfield  {author} {\bibinfo {author} {\bibfnamefont {Y.}~\bibnamefont
  {Ji}}, \bibinfo {author} {\bibfnamefont {X.}~\bibnamefont {Chen}}, \bibinfo
  {author} {\bibfnamefont {I.}~\bibnamefont {Polian}},\ and\ \bibinfo {author}
  {\bibfnamefont {Y.}~\bibnamefont {Ban}},\ }\bibfield  {title} {\bibinfo
  {title} {Algorithm-oriented qubit mapping for variational quantum
  algorithms},\ }\href@noop {} {\bibfield  {journal} {\bibinfo  {journal}
  {Physical Review Applied}\ }\textbf {\bibinfo {volume} {23}},\ \bibinfo
  {pages} {034022} (\bibinfo {year} {2025})}\BibitemShut {NoStop}%
\bibitem [{\citenamefont {Bittel}\ and\ \citenamefont
  {Kliesch}(2021)}]{bittel2021training}%
  \BibitemOpen
  \bibfield  {author} {\bibinfo {author} {\bibfnamefont {L.}~\bibnamefont
  {Bittel}}\ and\ \bibinfo {author} {\bibfnamefont {M.}~\bibnamefont
  {Kliesch}},\ }\bibfield  {title} {\bibinfo {title} {Training variational
  quantum algorithms is np-hard},\ }\href@noop {} {\bibfield  {journal}
  {\bibinfo  {journal} {Physical Review Letters}\ }\textbf {\bibinfo {volume}
  {127}},\ \bibinfo {pages} {120502} (\bibinfo {year} {2021})}\BibitemShut
  {NoStop}%
\bibitem [{\citenamefont {Wang}\ \emph {et~al.}(2021)\citenamefont {Wang},
  \citenamefont {Fontana}, \citenamefont {Cerezo}, \citenamefont {Sharma},
  \citenamefont {Sone}, \citenamefont {Cincio},\ and\ \citenamefont
  {Coles}}]{wang2021noise}%
  \BibitemOpen
  \bibfield  {author} {\bibinfo {author} {\bibfnamefont {S.}~\bibnamefont
  {Wang}}, \bibinfo {author} {\bibfnamefont {E.}~\bibnamefont {Fontana}},
  \bibinfo {author} {\bibfnamefont {M.}~\bibnamefont {Cerezo}}, \bibinfo
  {author} {\bibfnamefont {K.}~\bibnamefont {Sharma}}, \bibinfo {author}
  {\bibfnamefont {A.}~\bibnamefont {Sone}}, \bibinfo {author} {\bibfnamefont
  {L.}~\bibnamefont {Cincio}},\ and\ \bibinfo {author} {\bibfnamefont {P.~J.}\
  \bibnamefont {Coles}},\ }\bibfield  {title} {\bibinfo {title} {Noise-induced
  barren plateaus in variational quantum algorithms},\ }\href@noop {}
  {\bibfield  {journal} {\bibinfo  {journal} {Nature communications}\ }\textbf
  {\bibinfo {volume} {12}},\ \bibinfo {pages} {1} (\bibinfo {year}
  {2021})}\BibitemShut {NoStop}%
\bibitem [{\citenamefont {Ferreira-Martins}\ \emph {et~al.}(2025)\citenamefont
  {Ferreira-Martins}, \citenamefont {Farias}, \citenamefont {Camilo},
  \citenamefont {Maciel}, \citenamefont {Tosta}, \citenamefont {Lin},
  \citenamefont {Alhajri}, \citenamefont {Haug},\ and\ \citenamefont
  {Aolita}}]{ferreira2025variational}%
  \BibitemOpen
  \bibfield  {author} {\bibinfo {author} {\bibfnamefont {A.~J.}\ \bibnamefont
  {Ferreira-Martins}}, \bibinfo {author} {\bibfnamefont {R.}~\bibnamefont
  {Farias}}, \bibinfo {author} {\bibfnamefont {G.}~\bibnamefont {Camilo}},
  \bibinfo {author} {\bibfnamefont {T.~O.}\ \bibnamefont {Maciel}}, \bibinfo
  {author} {\bibfnamefont {A.}~\bibnamefont {Tosta}}, \bibinfo {author}
  {\bibfnamefont {R.}~\bibnamefont {Lin}}, \bibinfo {author} {\bibfnamefont
  {A.}~\bibnamefont {Alhajri}}, \bibinfo {author} {\bibfnamefont
  {T.}~\bibnamefont {Haug}},\ and\ \bibinfo {author} {\bibfnamefont
  {L.}~\bibnamefont {Aolita}},\ }\bibfield  {title} {\bibinfo {title}
  {Variational quantum algorithms with exact geodesic transport},\ }\href@noop
  {} {\bibfield  {journal} {\bibinfo  {journal} {arXiv preprint
  arXiv:2506.17395}\ } (\bibinfo {year} {2025})}\BibitemShut {NoStop}%
\bibitem [{\citenamefont {Du}\ \emph {et~al.}(2022)\citenamefont {Du},
  \citenamefont {Tu}, \citenamefont {Yuan},\ and\ \citenamefont
  {Tao}}]{du2022efficient}%
  \BibitemOpen
  \bibfield  {author} {\bibinfo {author} {\bibfnamefont {Y.}~\bibnamefont
  {Du}}, \bibinfo {author} {\bibfnamefont {Z.}~\bibnamefont {Tu}}, \bibinfo
  {author} {\bibfnamefont {X.}~\bibnamefont {Yuan}},\ and\ \bibinfo {author}
  {\bibfnamefont {D.}~\bibnamefont {Tao}},\ }\bibfield  {title} {\bibinfo
  {title} {Efficient measure for the expressivity of variational quantum
  algorithms},\ }\href@noop {} {\bibfield  {journal} {\bibinfo  {journal}
  {Physical Review Letters}\ }\textbf {\bibinfo {volume} {128}},\ \bibinfo
  {pages} {080506} (\bibinfo {year} {2022})}\BibitemShut {NoStop}%
\bibitem [{\citenamefont {Zhu}\ \emph {et~al.}(1997)\citenamefont {Zhu},
  \citenamefont {Byrd}, \citenamefont {Lu},\ and\ \citenamefont
  {Nocedal}}]{zhu1997algorithm}%
  \BibitemOpen
  \bibfield  {author} {\bibinfo {author} {\bibfnamefont {C.}~\bibnamefont
  {Zhu}}, \bibinfo {author} {\bibfnamefont {R.~H.}\ \bibnamefont {Byrd}},
  \bibinfo {author} {\bibfnamefont {P.}~\bibnamefont {Lu}},\ and\ \bibinfo
  {author} {\bibfnamefont {J.}~\bibnamefont {Nocedal}},\ }\bibfield  {title}
  {\bibinfo {title} {Algorithm 778: L-bfgs-b: Fortran subroutines for
  large-scale bound-constrained optimization},\ }\href@noop {} {\bibfield
  {journal} {\bibinfo  {journal} {ACM Transactions on mathematical software
  (TOMS)}\ }\textbf {\bibinfo {volume} {23}},\ \bibinfo {pages} {550} (\bibinfo
  {year} {1997})}\BibitemShut {NoStop}%
\bibitem [{\citenamefont {Shen}\ \emph
  {et~al.}(2025{\natexlab{c}})\citenamefont {Shen}, \citenamefont {Chen},
  \citenamefont {Yang},\ and\ \citenamefont {Lee}}]{shen2025observation}%
  \BibitemOpen
  \bibfield  {author} {\bibinfo {author} {\bibfnamefont {R.}~\bibnamefont
  {Shen}}, \bibinfo {author} {\bibfnamefont {T.}~\bibnamefont {Chen}}, \bibinfo
  {author} {\bibfnamefont {B.}~\bibnamefont {Yang}},\ and\ \bibinfo {author}
  {\bibfnamefont {C.~H.}\ \bibnamefont {Lee}},\ }\bibfield  {title} {\bibinfo
  {title} {Observation of the non-hermitian skin effect and fermi skin on a
  digital quantum computer},\ }\href@noop {} {\bibfield  {journal} {\bibinfo
  {journal} {Nature Communications}\ }\textbf {\bibinfo {volume} {16}},\
  \bibinfo {pages} {1340} (\bibinfo {year} {2025}{\natexlab{c}})}\BibitemShut
  {NoStop}%
\bibitem [{\citenamefont {Sun}\ \emph {et~al.}(2021)\citenamefont {Sun},
  \citenamefont {Motta}, \citenamefont {Tazhigulov}, \citenamefont {Tan},
  \citenamefont {Chan},\ and\ \citenamefont {Minnich}}]{sun2021quantum}%
  \BibitemOpen
  \bibfield  {author} {\bibinfo {author} {\bibfnamefont {S.-N.}\ \bibnamefont
  {Sun}}, \bibinfo {author} {\bibfnamefont {M.}~\bibnamefont {Motta}}, \bibinfo
  {author} {\bibfnamefont {R.~N.}\ \bibnamefont {Tazhigulov}}, \bibinfo
  {author} {\bibfnamefont {A.~T.}\ \bibnamefont {Tan}}, \bibinfo {author}
  {\bibfnamefont {G.~K.-L.}\ \bibnamefont {Chan}},\ and\ \bibinfo {author}
  {\bibfnamefont {A.~J.}\ \bibnamefont {Minnich}},\ }\bibfield  {title}
  {\bibinfo {title} {Quantum computation of finite-temperature static and
  dynamical properties of spin systems using quantum imaginary time
  evolution},\ }\href@noop {} {\bibfield  {journal} {\bibinfo  {journal} {PRX
  Quantum}\ }\textbf {\bibinfo {volume} {2}},\ \bibinfo {pages} {010317}
  (\bibinfo {year} {2021})}\BibitemShut {NoStop}%
\bibitem [{\citenamefont {Kamakari}\ \emph {et~al.}(2022)\citenamefont
  {Kamakari}, \citenamefont {Sun}, \citenamefont {Motta},\ and\ \citenamefont
  {Minnich}}]{kamakari2022digital}%
  \BibitemOpen
  \bibfield  {author} {\bibinfo {author} {\bibfnamefont {H.}~\bibnamefont
  {Kamakari}}, \bibinfo {author} {\bibfnamefont {S.-N.}\ \bibnamefont {Sun}},
  \bibinfo {author} {\bibfnamefont {M.}~\bibnamefont {Motta}},\ and\ \bibinfo
  {author} {\bibfnamefont {A.~J.}\ \bibnamefont {Minnich}},\ }\bibfield
  {title} {\bibinfo {title} {Digital quantum simulation of open quantum systems
  using quantum imaginary--time evolution},\ }\href@noop {} {\bibfield
  {journal} {\bibinfo  {journal} {PRX Quantum}\ }\textbf {\bibinfo {volume}
  {3}},\ \bibinfo {pages} {010320} (\bibinfo {year} {2022})}\BibitemShut
  {NoStop}%
\bibitem [{\citenamefont {McArdle}\ \emph {et~al.}(2019)\citenamefont
  {McArdle}, \citenamefont {Jones}, \citenamefont {Endo}, \citenamefont {Li},
  \citenamefont {Benjamin},\ and\ \citenamefont
  {Yuan}}]{mcardle2019variational}%
  \BibitemOpen
  \bibfield  {author} {\bibinfo {author} {\bibfnamefont {S.}~\bibnamefont
  {McArdle}}, \bibinfo {author} {\bibfnamefont {T.}~\bibnamefont {Jones}},
  \bibinfo {author} {\bibfnamefont {S.}~\bibnamefont {Endo}}, \bibinfo {author}
  {\bibfnamefont {Y.}~\bibnamefont {Li}}, \bibinfo {author} {\bibfnamefont
  {S.~C.}\ \bibnamefont {Benjamin}},\ and\ \bibinfo {author} {\bibfnamefont
  {X.}~\bibnamefont {Yuan}},\ }\bibfield  {title} {\bibinfo {title}
  {Variational ansatz-based quantum simulation of imaginary time evolution},\
  }\href@noop {} {\bibfield  {journal} {\bibinfo  {journal} {npj Quantum
  Information}\ }\textbf {\bibinfo {volume} {5}},\ \bibinfo {pages} {75}
  (\bibinfo {year} {2019})}\BibitemShut {NoStop}%
\bibitem [{\citenamefont {Motta}\ \emph {et~al.}(2020)\citenamefont {Motta},
  \citenamefont {Sun}, \citenamefont {Tan}, \citenamefont {O’Rourke},
  \citenamefont {Ye}, \citenamefont {Minnich}, \citenamefont {Brand{\~a}o},\
  and\ \citenamefont {Chan}}]{motta2020determining}%
  \BibitemOpen
  \bibfield  {author} {\bibinfo {author} {\bibfnamefont {M.}~\bibnamefont
  {Motta}}, \bibinfo {author} {\bibfnamefont {C.}~\bibnamefont {Sun}}, \bibinfo
  {author} {\bibfnamefont {A.~T.}\ \bibnamefont {Tan}}, \bibinfo {author}
  {\bibfnamefont {M.~J.}\ \bibnamefont {O’Rourke}}, \bibinfo {author}
  {\bibfnamefont {E.}~\bibnamefont {Ye}}, \bibinfo {author} {\bibfnamefont
  {A.~J.}\ \bibnamefont {Minnich}}, \bibinfo {author} {\bibfnamefont {F.~G.}\
  \bibnamefont {Brand{\~a}o}},\ and\ \bibinfo {author} {\bibfnamefont
  {G.~K.-L.}\ \bibnamefont {Chan}},\ }\bibfield  {title} {\bibinfo {title}
  {Determining eigenstates and thermal states on a quantum computer using
  quantum imaginary time evolution},\ }\href@noop {} {\bibfield  {journal}
  {\bibinfo  {journal} {Nature Physics}\ }\textbf {\bibinfo {volume} {16}},\
  \bibinfo {pages} {205} (\bibinfo {year} {2020})}\BibitemShut {NoStop}%
\bibitem [{\citenamefont {Shibata}\ and\ \citenamefont
  {Katsura}(2019)}]{shibata2019dissipative}%
  \BibitemOpen
  \bibfield  {author} {\bibinfo {author} {\bibfnamefont {N.}~\bibnamefont
  {Shibata}}\ and\ \bibinfo {author} {\bibfnamefont {H.}~\bibnamefont
  {Katsura}},\ }\bibfield  {title} {\bibinfo {title} {Dissipative quantum ising
  chain as a non-hermitian ashkin-teller model},\ }\href@noop {} {\bibfield
  {journal} {\bibinfo  {journal} {Physical Review B}\ }\textbf {\bibinfo
  {volume} {99}},\ \bibinfo {pages} {224432} (\bibinfo {year}
  {2019})}\BibitemShut {NoStop}%
\bibitem [{\citenamefont {Shen}\ \emph {et~al.}(2024)\citenamefont {Shen},
  \citenamefont {Qin}, \citenamefont {Desaules}, \citenamefont {Papi{\'c}},\
  and\ \citenamefont {Lee}}]{shen2024enhanced}%
  \BibitemOpen
  \bibfield  {author} {\bibinfo {author} {\bibfnamefont {R.}~\bibnamefont
  {Shen}}, \bibinfo {author} {\bibfnamefont {F.}~\bibnamefont {Qin}}, \bibinfo
  {author} {\bibfnamefont {J.-Y.}\ \bibnamefont {Desaules}}, \bibinfo {author}
  {\bibfnamefont {Z.}~\bibnamefont {Papi{\'c}}},\ and\ \bibinfo {author}
  {\bibfnamefont {C.~H.}\ \bibnamefont {Lee}},\ }\bibfield  {title} {\bibinfo
  {title} {Enhanced many-body quantum scars from the non-hermitian fock skin
  effect},\ }\href@noop {} {\bibfield  {journal} {\bibinfo  {journal} {Physical
  Review Letters}\ }\textbf {\bibinfo {volume} {133}},\ \bibinfo {pages}
  {216601} (\bibinfo {year} {2024})}\BibitemShut {NoStop}%
\bibitem [{\citenamefont {Xue}\ and\ \citenamefont
  {Lee}(2024)}]{xue2024topologically}%
  \BibitemOpen
  \bibfield  {author} {\bibinfo {author} {\bibfnamefont {W.-T.}\ \bibnamefont
  {Xue}}\ and\ \bibinfo {author} {\bibfnamefont {C.~H.}\ \bibnamefont {Lee}},\
  }\bibfield  {title} {\bibinfo {title} {Topologically protected negative
  entanglement},\ }\href@noop {} {\bibfield  {journal} {\bibinfo  {journal}
  {arXiv preprint arXiv:2403.03259}\ } (\bibinfo {year} {2024})}\BibitemShut
  {NoStop}%
\bibitem [{\citenamefont {Yang}\ \emph {et~al.}(2022)\citenamefont {Yang},
  \citenamefont {Raymond},\ and\ \citenamefont {Uno}}]{yang2022efficient}%
  \BibitemOpen
  \bibfield  {author} {\bibinfo {author} {\bibfnamefont {B.}~\bibnamefont
  {Yang}}, \bibinfo {author} {\bibfnamefont {R.}~\bibnamefont {Raymond}},\ and\
  \bibinfo {author} {\bibfnamefont {S.}~\bibnamefont {Uno}},\ }\bibfield
  {title} {\bibinfo {title} {Efficient quantum readout-error mitigation for
  sparse measurement outcomes of near-term quantum devices},\ }\href@noop {}
  {\bibfield  {journal} {\bibinfo  {journal} {Physical Review A}\ }\textbf
  {\bibinfo {volume} {106}},\ \bibinfo {pages} {012423} (\bibinfo {year}
  {2022})}\BibitemShut {NoStop}%
\bibitem [{\citenamefont {Nachman}\ \emph {et~al.}(2020)\citenamefont
  {Nachman}, \citenamefont {Urbanek}, \citenamefont {de~Jong},\ and\
  \citenamefont {Bauer}}]{nachman2020unfolding}%
  \BibitemOpen
  \bibfield  {author} {\bibinfo {author} {\bibfnamefont {B.}~\bibnamefont
  {Nachman}}, \bibinfo {author} {\bibfnamefont {M.}~\bibnamefont {Urbanek}},
  \bibinfo {author} {\bibfnamefont {W.~A.}\ \bibnamefont {de~Jong}},\ and\
  \bibinfo {author} {\bibfnamefont {C.~W.}\ \bibnamefont {Bauer}},\ }\bibfield
  {title} {\bibinfo {title} {Unfolding quantum computer readout noise},\
  }\href@noop {} {\bibfield  {journal} {\bibinfo  {journal} {npj Quantum
  Information}\ }\textbf {\bibinfo {volume} {6}},\ \bibinfo {pages} {84}
  (\bibinfo {year} {2020})}\BibitemShut {NoStop}%
\bibitem [{\citenamefont {Liu}\ \emph {et~al.}(2022)\citenamefont {Liu},
  \citenamefont {Shtengel}, \citenamefont {Smith},\ and\ \citenamefont
  {Pollmann}}]{liu2022methods}%
  \BibitemOpen
  \bibfield  {author} {\bibinfo {author} {\bibfnamefont {Y.-J.}\ \bibnamefont
  {Liu}}, \bibinfo {author} {\bibfnamefont {K.}~\bibnamefont {Shtengel}},
  \bibinfo {author} {\bibfnamefont {A.}~\bibnamefont {Smith}},\ and\ \bibinfo
  {author} {\bibfnamefont {F.}~\bibnamefont {Pollmann}},\ }\bibfield  {title}
  {\bibinfo {title} {Methods for simulating string-net states and anyons on a
  digital quantum computer},\ }\href@noop {} {\bibfield  {journal} {\bibinfo
  {journal} {PRX Quantum}\ }\textbf {\bibinfo {volume} {3}},\ \bibinfo {pages}
  {040315} (\bibinfo {year} {2022})}\BibitemShut {NoStop}%
\bibitem [{\citenamefont {Xu}\ \emph {et~al.}(2024)\citenamefont {Xu},
  \citenamefont {Sun}, \citenamefont {Wang}, \citenamefont {Li}, \citenamefont
  {Zhu}, \citenamefont {Dong}, \citenamefont {Deng}, \citenamefont {Zhang},
  \citenamefont {Chen}, \citenamefont {Wu} \emph {et~al.}}]{xu2024non}%
  \BibitemOpen
  \bibfield  {author} {\bibinfo {author} {\bibfnamefont {S.}~\bibnamefont
  {Xu}}, \bibinfo {author} {\bibfnamefont {Z.-Z.}\ \bibnamefont {Sun}},
  \bibinfo {author} {\bibfnamefont {K.}~\bibnamefont {Wang}}, \bibinfo {author}
  {\bibfnamefont {H.}~\bibnamefont {Li}}, \bibinfo {author} {\bibfnamefont
  {Z.}~\bibnamefont {Zhu}}, \bibinfo {author} {\bibfnamefont {H.}~\bibnamefont
  {Dong}}, \bibinfo {author} {\bibfnamefont {J.}~\bibnamefont {Deng}}, \bibinfo
  {author} {\bibfnamefont {X.}~\bibnamefont {Zhang}}, \bibinfo {author}
  {\bibfnamefont {J.}~\bibnamefont {Chen}}, \bibinfo {author} {\bibfnamefont
  {Y.}~\bibnamefont {Wu}}, \emph {et~al.},\ }\bibfield  {title} {\bibinfo
  {title} {Non-abelian braiding of fibonacci anyons with a superconducting
  processor},\ }\href@noop {} {\bibfield  {journal} {\bibinfo  {journal}
  {Nature Physics}\ }\textbf {\bibinfo {volume} {20}},\ \bibinfo {pages} {1469}
  (\bibinfo {year} {2024})}\BibitemShut {NoStop}%
\bibitem [{\citenamefont {Xu}\ \emph {et~al.}(2023)\citenamefont {Xu},
  \citenamefont {Sun}, \citenamefont {Wang}, \citenamefont {Xiang},
  \citenamefont {Bao}, \citenamefont {Zhu}, \citenamefont {Shen}, \citenamefont
  {Song}, \citenamefont {Zhang}, \citenamefont {Ren} \emph
  {et~al.}}]{xu2023digital}%
  \BibitemOpen
  \bibfield  {author} {\bibinfo {author} {\bibfnamefont {S.}~\bibnamefont
  {Xu}}, \bibinfo {author} {\bibfnamefont {Z.-Z.}\ \bibnamefont {Sun}},
  \bibinfo {author} {\bibfnamefont {K.}~\bibnamefont {Wang}}, \bibinfo {author}
  {\bibfnamefont {L.}~\bibnamefont {Xiang}}, \bibinfo {author} {\bibfnamefont
  {Z.}~\bibnamefont {Bao}}, \bibinfo {author} {\bibfnamefont {Z.}~\bibnamefont
  {Zhu}}, \bibinfo {author} {\bibfnamefont {F.}~\bibnamefont {Shen}}, \bibinfo
  {author} {\bibfnamefont {Z.}~\bibnamefont {Song}}, \bibinfo {author}
  {\bibfnamefont {P.}~\bibnamefont {Zhang}}, \bibinfo {author} {\bibfnamefont
  {W.}~\bibnamefont {Ren}}, \emph {et~al.},\ }\bibfield  {title} {\bibinfo
  {title} {Digital simulation of projective non-abelian anyons with 68
  superconducting qubits},\ }\href@noop {} {\bibfield  {journal} {\bibinfo
  {journal} {Chinese Physics Letters}\ }\textbf {\bibinfo {volume} {40}},\
  \bibinfo {pages} {060301} (\bibinfo {year} {2023})}\BibitemShut {NoStop}%
\bibitem [{\citenamefont {Lanyon}\ \emph {et~al.}(2010)\citenamefont {Lanyon},
  \citenamefont {Whitfield}, \citenamefont {Gillett}, \citenamefont {Goggin},
  \citenamefont {Almeida}, \citenamefont {Kassal}, \citenamefont {Biamonte},
  \citenamefont {Mohseni}, \citenamefont {Powell}, \citenamefont {Barbieri}
  \emph {et~al.}}]{lanyon2010towards}%
  \BibitemOpen
  \bibfield  {author} {\bibinfo {author} {\bibfnamefont {B.~P.}\ \bibnamefont
  {Lanyon}}, \bibinfo {author} {\bibfnamefont {J.~D.}\ \bibnamefont
  {Whitfield}}, \bibinfo {author} {\bibfnamefont {G.~G.}\ \bibnamefont
  {Gillett}}, \bibinfo {author} {\bibfnamefont {M.~E.}\ \bibnamefont {Goggin}},
  \bibinfo {author} {\bibfnamefont {M.~P.}\ \bibnamefont {Almeida}}, \bibinfo
  {author} {\bibfnamefont {I.}~\bibnamefont {Kassal}}, \bibinfo {author}
  {\bibfnamefont {J.~D.}\ \bibnamefont {Biamonte}}, \bibinfo {author}
  {\bibfnamefont {M.}~\bibnamefont {Mohseni}}, \bibinfo {author} {\bibfnamefont
  {B.~J.}\ \bibnamefont {Powell}}, \bibinfo {author} {\bibfnamefont
  {M.}~\bibnamefont {Barbieri}}, \emph {et~al.},\ }\bibfield  {title} {\bibinfo
  {title} {Towards quantum chemistry on a quantum computer},\ }\href@noop {}
  {\bibfield  {journal} {\bibinfo  {journal} {Nature chemistry}\ }\textbf
  {\bibinfo {volume} {2}},\ \bibinfo {pages} {106} (\bibinfo {year}
  {2010})}\BibitemShut {NoStop}%
\bibitem [{\citenamefont {Wecker}\ \emph {et~al.}(2014)\citenamefont {Wecker},
  \citenamefont {Bauer}, \citenamefont {Clark}, \citenamefont {Hastings},\ and\
  \citenamefont {Troyer}}]{wecker2014gate}%
  \BibitemOpen
  \bibfield  {author} {\bibinfo {author} {\bibfnamefont {D.}~\bibnamefont
  {Wecker}}, \bibinfo {author} {\bibfnamefont {B.}~\bibnamefont {Bauer}},
  \bibinfo {author} {\bibfnamefont {B.~K.}\ \bibnamefont {Clark}}, \bibinfo
  {author} {\bibfnamefont {M.~B.}\ \bibnamefont {Hastings}},\ and\ \bibinfo
  {author} {\bibfnamefont {M.}~\bibnamefont {Troyer}},\ }\bibfield  {title}
  {\bibinfo {title} {Gate-count estimates for performing quantum chemistry on
  small quantum computers},\ }\href@noop {} {\bibfield  {journal} {\bibinfo
  {journal} {Physical Review A}\ }\textbf {\bibinfo {volume} {90}},\ \bibinfo
  {pages} {022305} (\bibinfo {year} {2014})}\BibitemShut {NoStop}%
\bibitem [{\citenamefont {Kassal}\ \emph {et~al.}(2011)\citenamefont {Kassal},
  \citenamefont {Whitfield}, \citenamefont {Perdomo-Ortiz}, \citenamefont
  {Yung},\ and\ \citenamefont {Aspuru-Guzik}}]{kassal2011simulating}%
  \BibitemOpen
  \bibfield  {author} {\bibinfo {author} {\bibfnamefont {I.}~\bibnamefont
  {Kassal}}, \bibinfo {author} {\bibfnamefont {J.~D.}\ \bibnamefont
  {Whitfield}}, \bibinfo {author} {\bibfnamefont {A.}~\bibnamefont
  {Perdomo-Ortiz}}, \bibinfo {author} {\bibfnamefont {M.-H.}\ \bibnamefont
  {Yung}},\ and\ \bibinfo {author} {\bibfnamefont {A.}~\bibnamefont
  {Aspuru-Guzik}},\ }\bibfield  {title} {\bibinfo {title} {Simulating chemistry
  using quantum computers},\ }\href@noop {} {\bibfield  {journal} {\bibinfo
  {journal} {Annual review of physical chemistry}\ }\textbf {\bibinfo {volume}
  {62}},\ \bibinfo {pages} {185} (\bibinfo {year} {2011})}\BibitemShut
  {NoStop}%
\bibitem [{\citenamefont {O’Brien}\ \emph {et~al.}(2019)\citenamefont
  {O’Brien}, \citenamefont {Senjean}, \citenamefont {Sagastizabal},
  \citenamefont {Bonet-Monroig}, \citenamefont {Dutkiewicz}, \citenamefont
  {Buda}, \citenamefont {DiCarlo},\ and\ \citenamefont
  {Visscher}}]{o2019calculating}%
  \BibitemOpen
  \bibfield  {author} {\bibinfo {author} {\bibfnamefont {T.~E.}\ \bibnamefont
  {O’Brien}}, \bibinfo {author} {\bibfnamefont {B.}~\bibnamefont {Senjean}},
  \bibinfo {author} {\bibfnamefont {R.}~\bibnamefont {Sagastizabal}}, \bibinfo
  {author} {\bibfnamefont {X.}~\bibnamefont {Bonet-Monroig}}, \bibinfo {author}
  {\bibfnamefont {A.}~\bibnamefont {Dutkiewicz}}, \bibinfo {author}
  {\bibfnamefont {F.}~\bibnamefont {Buda}}, \bibinfo {author} {\bibfnamefont
  {L.}~\bibnamefont {DiCarlo}},\ and\ \bibinfo {author} {\bibfnamefont
  {L.}~\bibnamefont {Visscher}},\ }\bibfield  {title} {\bibinfo {title}
  {Calculating energy derivatives for quantum chemistry on a quantum
  computer},\ }\href@noop {} {\bibfield  {journal} {\bibinfo  {journal} {npj
  Quantum Information}\ }\textbf {\bibinfo {volume} {5}},\ \bibinfo {pages}
  {113} (\bibinfo {year} {2019})}\BibitemShut {NoStop}%
\bibitem [{\citenamefont {Gaita-Ari{\~n}o}\ \emph {et~al.}(2019)\citenamefont
  {Gaita-Ari{\~n}o}, \citenamefont {Luis}, \citenamefont {Hill},\ and\
  \citenamefont {Coronado}}]{gaita2019molecular}%
  \BibitemOpen
  \bibfield  {author} {\bibinfo {author} {\bibfnamefont {A.}~\bibnamefont
  {Gaita-Ari{\~n}o}}, \bibinfo {author} {\bibfnamefont {F.}~\bibnamefont
  {Luis}}, \bibinfo {author} {\bibfnamefont {S.}~\bibnamefont {Hill}},\ and\
  \bibinfo {author} {\bibfnamefont {E.}~\bibnamefont {Coronado}},\ }\bibfield
  {title} {\bibinfo {title} {Molecular spins for quantum computation},\
  }\href@noop {} {\bibfield  {journal} {\bibinfo  {journal} {Nature chemistry}\
  }\textbf {\bibinfo {volume} {11}},\ \bibinfo {pages} {301} (\bibinfo {year}
  {2019})}\BibitemShut {NoStop}%
\bibitem [{\citenamefont {Ma}\ \emph {et~al.}(2020)\citenamefont {Ma},
  \citenamefont {Govoni},\ and\ \citenamefont {Galli}}]{ma2020quantum}%
  \BibitemOpen
  \bibfield  {author} {\bibinfo {author} {\bibfnamefont {H.}~\bibnamefont
  {Ma}}, \bibinfo {author} {\bibfnamefont {M.}~\bibnamefont {Govoni}},\ and\
  \bibinfo {author} {\bibfnamefont {G.}~\bibnamefont {Galli}},\ }\bibfield
  {title} {\bibinfo {title} {Quantum simulations of materials on near-term
  quantum computers},\ }\href@noop {} {\bibfield  {journal} {\bibinfo
  {journal} {npj Computational Materials}\ }\textbf {\bibinfo {volume} {6}},\
  \bibinfo {pages} {85} (\bibinfo {year} {2020})}\BibitemShut {NoStop}%
\bibitem [{\citenamefont {Bauer}\ \emph {et~al.}(2020)\citenamefont {Bauer},
  \citenamefont {Bravyi}, \citenamefont {Motta},\ and\ \citenamefont
  {Chan}}]{bauer2020quantum}%
  \BibitemOpen
  \bibfield  {author} {\bibinfo {author} {\bibfnamefont {B.}~\bibnamefont
  {Bauer}}, \bibinfo {author} {\bibfnamefont {S.}~\bibnamefont {Bravyi}},
  \bibinfo {author} {\bibfnamefont {M.}~\bibnamefont {Motta}},\ and\ \bibinfo
  {author} {\bibfnamefont {G.~K.-L.}\ \bibnamefont {Chan}},\ }\bibfield
  {title} {\bibinfo {title} {Quantum algorithms for quantum chemistry and
  quantum materials science},\ }\href@noop {} {\bibfield  {journal} {\bibinfo
  {journal} {Chemical Reviews}\ }\textbf {\bibinfo {volume} {120}},\ \bibinfo
  {pages} {12685} (\bibinfo {year} {2020})}\BibitemShut {NoStop}%
\bibitem [{\citenamefont {de~Leon}\ \emph {et~al.}(2021)\citenamefont
  {de~Leon}, \citenamefont {Itoh}, \citenamefont {Kim}, \citenamefont {Mehta},
  \citenamefont {Northup}, \citenamefont {Paik}, \citenamefont {Palmer},
  \citenamefont {Samarth}, \citenamefont {Sangtawesin},\ and\ \citenamefont
  {Steuerman}}]{de2021materials}%
  \BibitemOpen
  \bibfield  {author} {\bibinfo {author} {\bibfnamefont {N.~P.}\ \bibnamefont
  {de~Leon}}, \bibinfo {author} {\bibfnamefont {K.~M.}\ \bibnamefont {Itoh}},
  \bibinfo {author} {\bibfnamefont {D.}~\bibnamefont {Kim}}, \bibinfo {author}
  {\bibfnamefont {K.~K.}\ \bibnamefont {Mehta}}, \bibinfo {author}
  {\bibfnamefont {T.~E.}\ \bibnamefont {Northup}}, \bibinfo {author}
  {\bibfnamefont {H.}~\bibnamefont {Paik}}, \bibinfo {author} {\bibfnamefont
  {B.}~\bibnamefont {Palmer}}, \bibinfo {author} {\bibfnamefont
  {N.}~\bibnamefont {Samarth}}, \bibinfo {author} {\bibfnamefont
  {S.}~\bibnamefont {Sangtawesin}},\ and\ \bibinfo {author} {\bibfnamefont
  {D.~W.}\ \bibnamefont {Steuerman}},\ }\bibfield  {title} {\bibinfo {title}
  {Materials challenges and opportunities for quantum computing hardware},\
  }\href@noop {} {\bibfield  {journal} {\bibinfo  {journal} {Science}\ }\textbf
  {\bibinfo {volume} {372}},\ \bibinfo {pages} {eabb2823} (\bibinfo {year}
  {2021})}\BibitemShut {NoStop}%
\bibitem [{\citenamefont {Lin}\ \emph {et~al.}(2021)\citenamefont {Lin},
  \citenamefont {Dilip}, \citenamefont {Green}, \citenamefont {Smith},\ and\
  \citenamefont {Pollmann}}]{lin2021real}%
  \BibitemOpen
  \bibfield  {author} {\bibinfo {author} {\bibfnamefont {S.-H.}\ \bibnamefont
  {Lin}}, \bibinfo {author} {\bibfnamefont {R.}~\bibnamefont {Dilip}}, \bibinfo
  {author} {\bibfnamefont {A.~G.}\ \bibnamefont {Green}}, \bibinfo {author}
  {\bibfnamefont {A.}~\bibnamefont {Smith}},\ and\ \bibinfo {author}
  {\bibfnamefont {F.}~\bibnamefont {Pollmann}},\ }\bibfield  {title} {\bibinfo
  {title} {Real-and imaginary-time evolution with compressed quantum
  circuits},\ }\href@noop {} {\bibfield  {journal} {\bibinfo  {journal} {PRX
  Quantum}\ }\textbf {\bibinfo {volume} {2}},\ \bibinfo {pages} {010342}
  (\bibinfo {year} {2021})}\BibitemShut {NoStop}%
\end{thebibliography}%

\onecolumngrid
\flushbottom
\newpage
\appendix
\setcounter{equation}{0}
\setcounter{figure}{0}
\setcounter{table}{0}
\setcounter{section}{0}
\renewcommand{\theequation}{S\arabic{equation}}
\renewcommand{\thefigure}{S\arabic{figure}}
\renewcommand{\thepage}{S\arabic{page}}
\newpage

\section{Ancilla-based approach}\label{apsec1}
\begin{figure}
	\centering
	\includegraphics[width=0.4\linewidth]{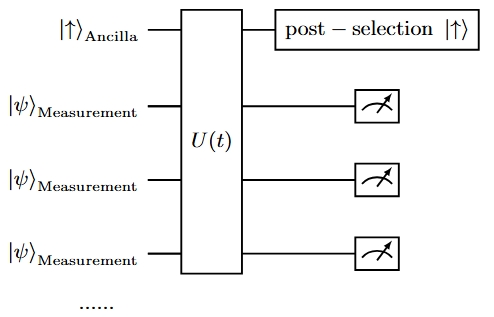}
	\caption{Quantum circuit implementation of nonunitary dynamics.
 $U(t)$ corresponds to the unitary embedding defined in Eq.\eqref{u}.  The first qubit serves as an ancilla qubit, requiring final post-selection on the state $\ket{\uparrow}$. Following this post-selection, the output state of measurement qubits is normalized to $e^{-itH}\ket{\psi}/\left\| e^{-itH}\ket{\psi}\right\|$, thereby ensuring the correct implementation of the nonunitary evolution.}
	\label{circuit}
\end{figure}

Here, we discuss the implementation of nonunitary operators, as introduced in Ref.~\cite{lin2021real,chen2022high}. To achieve this, we embed the nonunitary operator $U_{\rm H}(t)=e^{-itH}$, with $H$ denoting a non-Hermitian Hamiltonian, into a unitary operator as follows:
\begin{equation}\label{u}
	U(t)=\left[\begin{array}{cc}
		uU_{\rm  H}(t) & B \\                                     
		C & D
	\end{array}\right].
\end{equation}
Here, the normalization factor $u^{-2}$ corresponds to the maximum eigenvalue of $U_{\rm  H}^{\dagger}U_{\rm  H}$. The matrix $C$ is given by $C=\sqrt{I-u^{2}U_{\rm  H}^{\dagger}U_{\rm  H}}$, which is numerically determined. To determine the remaining unknown blocks $B$ and $D$,  we employ QR decomposition on the following ansatz matrix:
\begin{equation}
\left[\begin{array}{cc}
		uU_{\rm H}(t) & I\\                            
		C& I
	\end{array}\right],
\end{equation}
where $I$ is our ansatz identity matrix. Then, $B$ and $D$ are solved as follows
\begin{equation}
	U(t)G=\left[\begin{array}{cc}
		uU_{\rm H}(t) & I\\                                            
		C& I
	\end{array}\right],
\end{equation}
where $G$ is an upper triangular matrix obtained from QR decomposition. The implementation of this operator $U(t)$ is shown in FIG.\ref{circuit}, where this circuit consists of $N$ measurement qubits and one ancilla qubit. Post-selection on the ancilla state $\uparrow$ projects the system onto the desired subspace and gives the result from measurement qubits normalized as $e^{-itH}\ket{\psi}/\left\| e^{-itH}\ket{\psi}\right\|$.

\section{Variational circuits}\label{apvqa}
In our work, we employ variational circuits to perform quantum simulations. These circuits are optimized by minimizing the following cost function:
\begin{equation}
	C({\bm \theta, \bm\phi, \bm\lambda})=1-|\bra{\psi_{0}}V^{\dagger}U(t)\ket{\psi_{0}}|,
\end{equation}
where $V$ represents a parameterized circuit, and $U(t)$ is the target unitary defined in Eq.~\ref{u}. 

\begin{figure}
	\centering
	\includegraphics[width=0.5\linewidth]{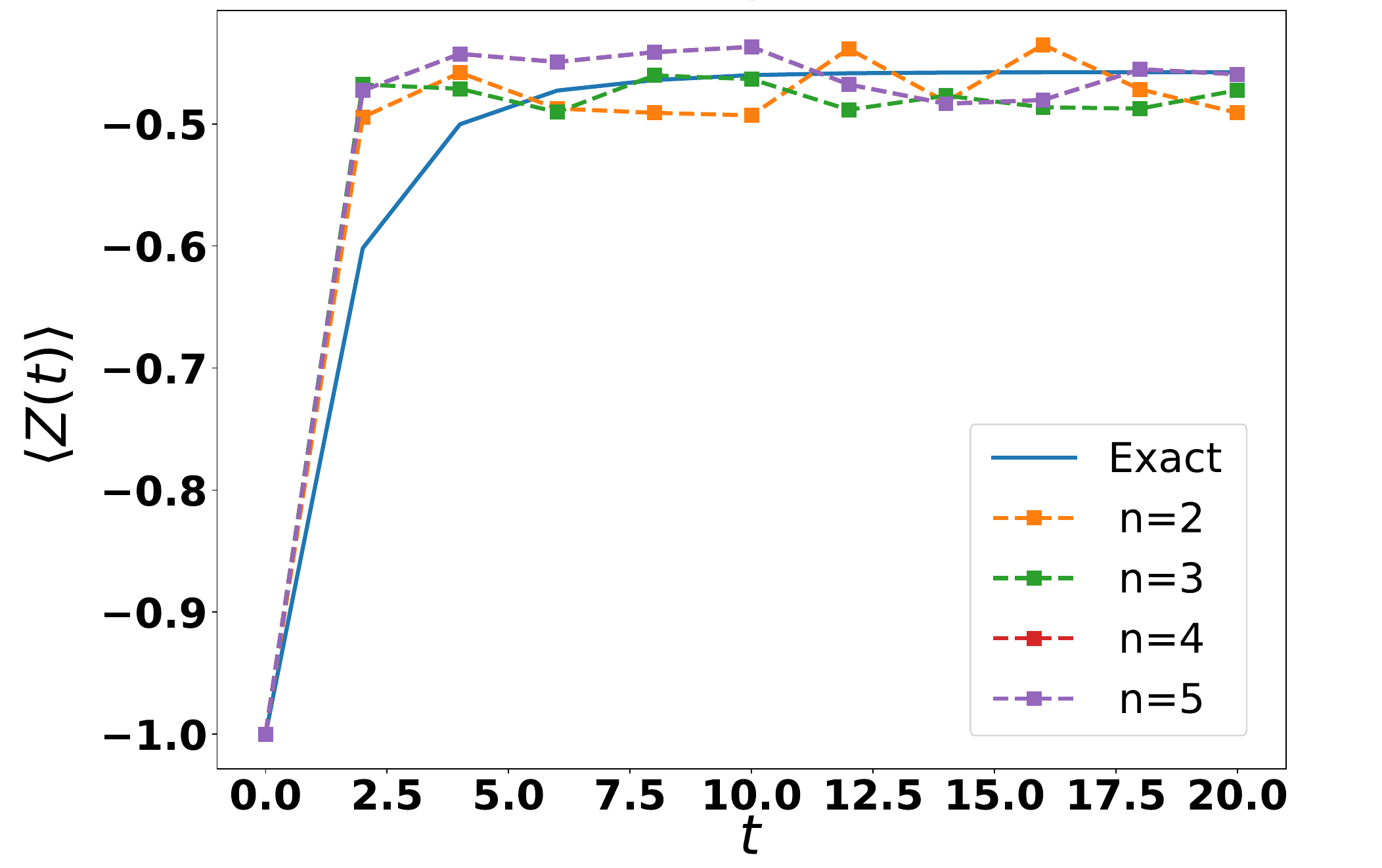}
\caption{
  Assessment of variational circuit fidelity.  We present the classical noiseless simulation of the dynamics $e^{-t{H}_{\rm TFI}}$ [Eq.~\ref{tfi}] and measure the evolution of $Z$ magnetization.  Here, the number of layers $n$ does not significantly impact the measured magnetization. While slight deviations from the exact result (blue curve) are observed due to sampling noise from the finite number of circuit executions, all dashed curves representing variational circuit outcomes closely align well with the exact result, confirming reliable convergence of numerical optimization. The initial state $\ket{\psi_{0}}=\ket{\psi_{p}}\ket{\psi_{a}}$ includes the measurement qubits initialized to  $\ket{\psi_{p}}=\ket{\downarrow\downarrow\downarrow\downarrow}$  and the ancilla state $\ket{\psi_{a}}=\ket{\uparrow}$. The parameterized circuit $V$ is depicted in FIG.\ref{fig:mps}, with each layer consisting of  $8$ CX gates. Other parameters are $J=1$, $\gamma=-0.5$ and $h_{x}=1.5$.}
	\label{fig:dynamics}
\end{figure}

Here, we execute the trained variational circuit on a classical simulator in the absence of noise and measure the $Z$ magnetization,  defined as $\langle Z(t) \rangle=\frac{1}{N}\sum_{i}\bra{\psi(t)}Z_{i}\ket{\psi(t)}$, with $N$ denoting the number of measurement qubits. The final normalized state is $\ket{\psi(t)}=e^{-it H_{\rm TFI}}\ket{\psi_{p}}/\left\| e^{-it H_{\rm TFI}}\ket{\psi_{p}}\right\| $. Notably, as shown in FIG.~\ref{fig:dynamics}, results indicate that the magnetization converges to a stable value over time, and the improvement from adding more layers is not significant. This suggests that the circuit with $n=2$ layers provides the optimal balance between fidelity and circuit depth. In the main text, we limit the ansatz circuit to a maximum of $n=5$ layers.

\begin{figure}
    \centering
    \includegraphics[width=0.5\linewidth]{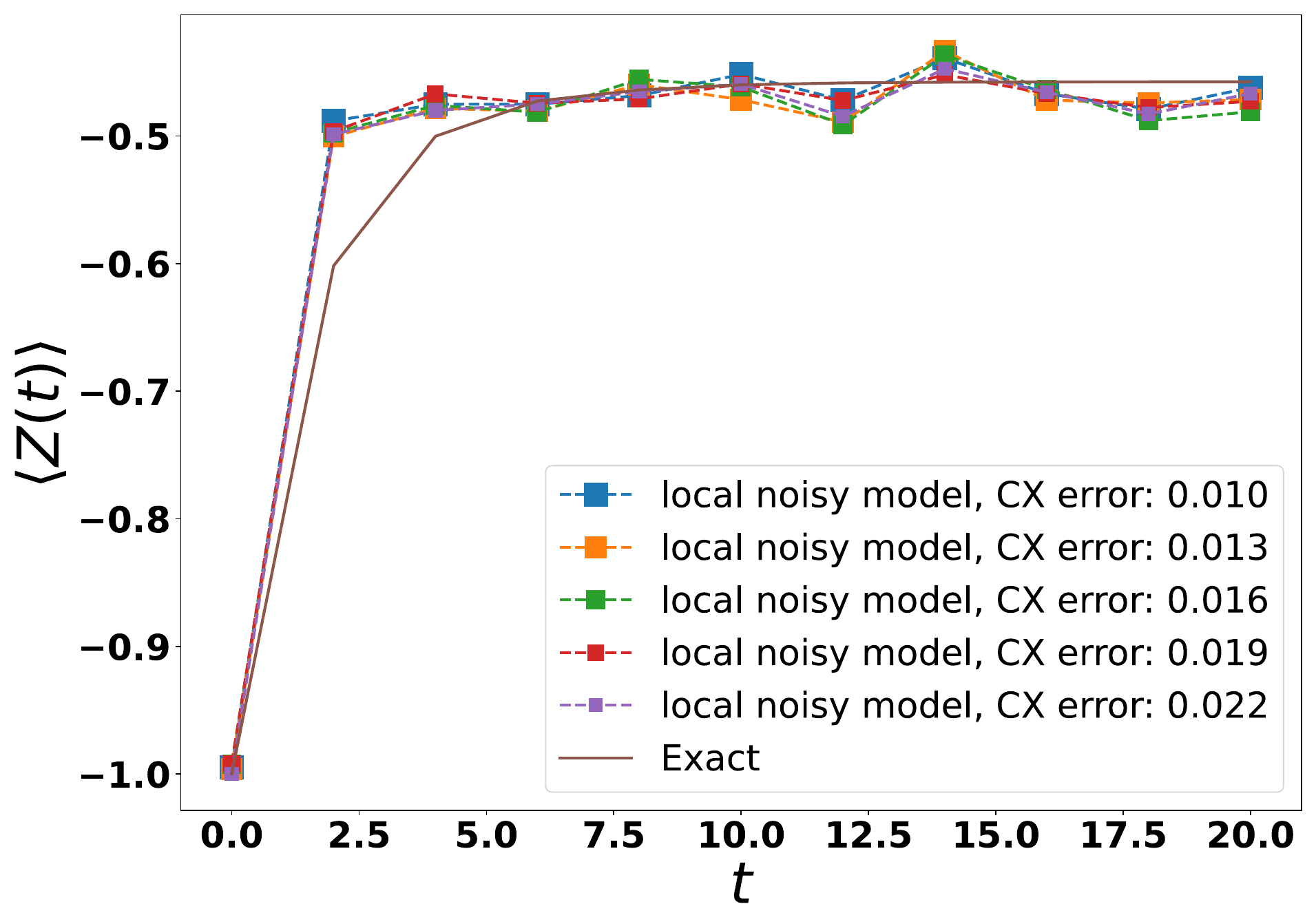}
    \caption{We present the numerical noisy simulation of the dynamics $e^{-t{H}_{\rm TFI}}$ [Eq.~\ref{tfi}] and measure the evolution of $Z$ magnetization. For noisy simulations, we apply our error mitigation under increasing levels of CX error rates. The mitigated results (dashed curves) closely match the exact outcome. The total circuit is built by $8$ layers.
    The initial state $\ket{\psi_{0}}=\ket{\psi_{p}}\ket{\psi_{a}}$ of $5$ qubits includes the initial state  $\ket{\psi_{p}}=\ket{\downarrow\downarrow\downarrow\downarrow}$  for measurement qubits and the ancilla state $\ket{\psi_{a}}=\ket{\uparrow}$. The parameterized circuit $V$ is depicted in FIG.\ref{fig:mps}, with each layer consisting of $8$ CX gates. Other parameters are $J=1$, $\gamma=-0.5$ and $h_{x}=1.5$.}
    \label{fig:enter-label}
\end{figure}

As shown in FIG.~\ref{fig:enter-label}, we implement our designed error mitigation method in a deeper circuit with $8$ layers, which can achieve better convergence in the cost function. Across all tested error models, the method consistently produces well-mitigated results that closely match exact results. However, for execution on hardware, deeper circuits can lead to longer processing times, potentially resulting in increased noise fluctuations due to temporal instability.

\begin{figure}
	\centering
	\includegraphics[width=0.99\linewidth]{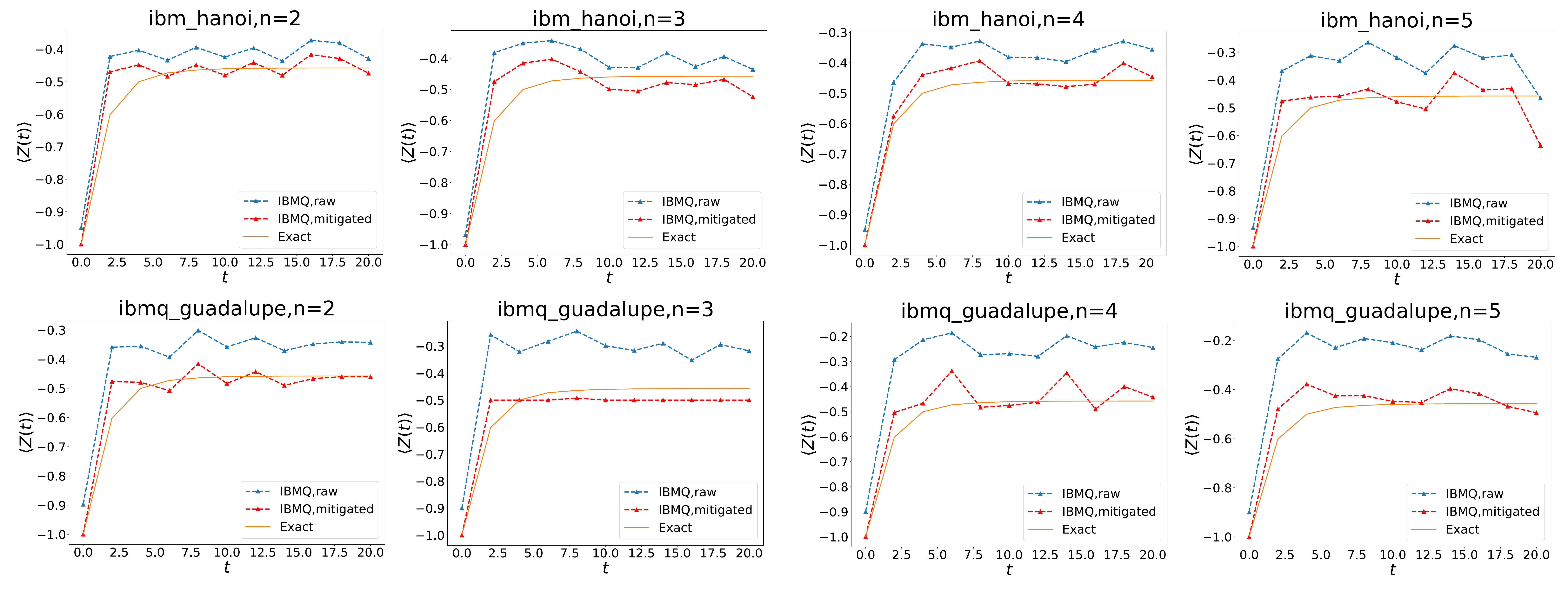}
	\caption{Additional quantum simulation results. We present quantum simulations of the dynamics $e^{-it\hat{H}_{\rm TFI}}$ governed by the model defined in Eq.\eqref{tfi}, The yellow curve represents the exact result, while the red and blue curves correspond to quantum simulations performed on IBM Q devices with and without our error mitigation method [Eq.~\ref{mit}], respectively. $n$ is the number of ansatz layers. The initial state is set to $\ket{\psi{0}}=\ket{\psi_{p}}\ket{\psi_{a}}$ of 5 qubits, where measurement qubits are initialized in the spin-down state 
$\ket{\psi_{p}}=\ket{\downarrow\downarrow\downarrow\downarrow}$ for measurement qubits, and the ancilla state is $\ket{\psi_{a}}=\ket{\uparrow}$. Across all circuit sizes, our error-mitigation method demonstrates great fidelity. Other parameters are $J=1$, $h_{x}=1.5$, and $\gamma=-0.5$. The number of shots is $32000$.  The IBM Q devices used in these simulations are ``ibmq Hanoi" and ``ibmq Guadalupe", with device conditions provided in Appendix.~\ref{ibmq}.}
	\label{fig:miterror}
\end{figure}

\section{Additional simulation results}

We here provide additional results simulated on two different IBM Q devices, each characterized by distinct noise profiles (see detailed device parameters in Appendix.\ref{ibmq}). As shown in FIG.\ref{fig:miterror}, for circuits with $n=2$ and $3$ layers, the error-mitigated outcomes from both devices exhibit excellent agreement with exact theoretical predictions. This strong agreement can be attributed to the minimal noise fluctuation during the execution of these relatively short circuits. As the circuit depth increases to $n=4$ layers, where gate error accumulation becomes more pronounced,  our enhanced mitigation approach continues to demonstrate robust performance, and noisy results consistently align with exact results. Even in the scenario of circuits with $n=5$ layers, despite some imperfections resulting from unstable noise characterization, our optimized method still achieves generally effective noise mitigation.

\section{IBM Q Quantum devices}\label{ibmq}

\begin{figure}[h]
	\centering
	\includegraphics[width=0.99\linewidth]{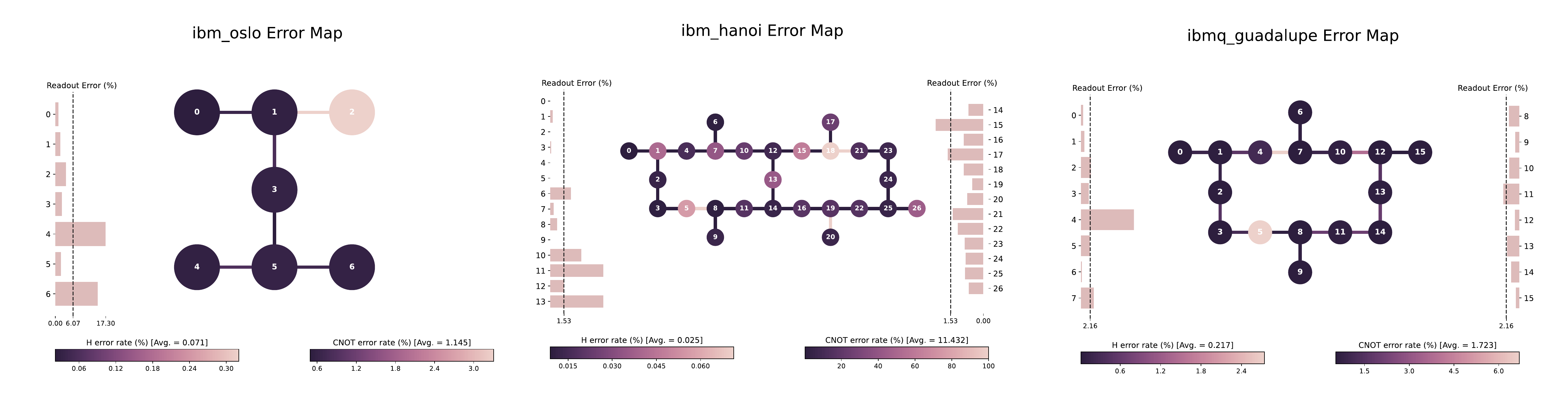}
	\caption{Qubit layout and calibration data for IBM Q devices.
 In our work, we deploy noisy quantum simulations on four IBM Q devices: {\bf IBM\_oslo}, {\bf IBM\_hanoi}, and {\bf IBM\_guadalupe}. The color of each edge is for the error rate of CX gates. The horizontal column represents the measurement error. Here, we select the following qubit chains for simulation: $[0,1,3,5,6]$ for  {\bf IBM\_oslo}, $[21,23,24,25,22]$ for {\bf IBM\_hanoi}, and $[9,8,11,14,13]$ for {\bf IBM\_guadalupe}, with mean CX error rates of $0.012$, $0.012$, and $0.015$  respectively. }
	\label{fig:errordata}
\end{figure}

FIG.~\ref{fig:errordata} presents detailed information on the IBM Q devices used in our simulations. To ensure optimal performance, we selected qubit chains with the best noise conditions for each device. Since our variational circuits are constructed using stacked CX gates, qubit selection was primarily based on minimizing CX gate errors. The selected qubit chains are: $[0,1,3,5,6]$ for  {\bf IBM\_oslo}, $[21,23,24,25,22]$ for {\bf IBM\_hanoi}, and $[9,8,11,14,13]$ for {\bf IBM\_guadalupe}, 
\section{Additional simulations}
\begin{figure}
	\centering
	\includegraphics[width=0.9\linewidth]{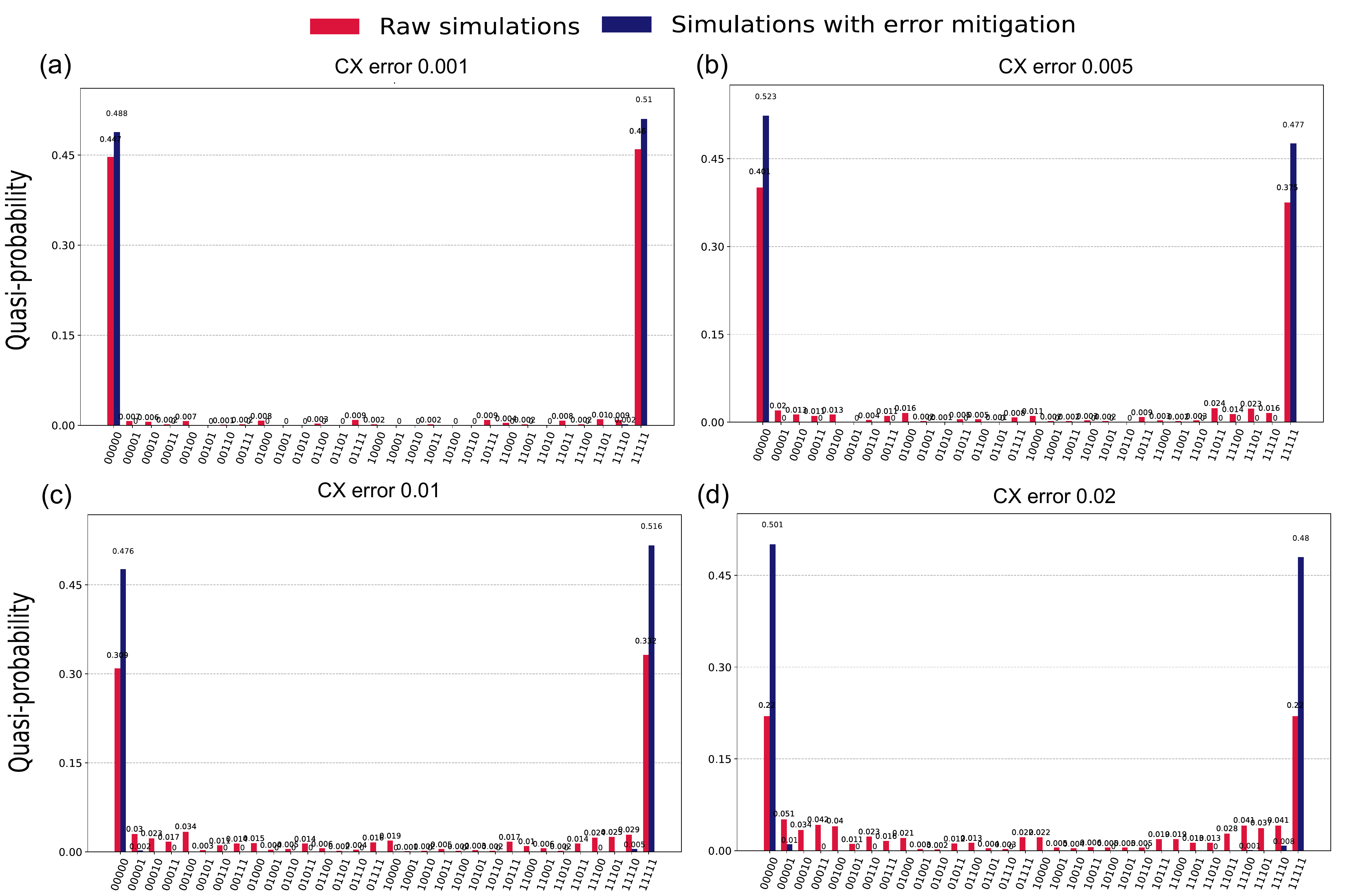}
	\caption{Noisy simulations of GHZ states. We employ a three-layer variational circuit to prepare the five-qubit GHZ state $\left(\ket{00000}+\ket{11111}\right)/\sqrt{2}$ on a local noisy simulator. We present the quasi-probability distributions of measurement outcomes for different two-qubit CX gate error rates. Panels (a–d) correspond to CX error rates of (a) 0.001, (b) 0.005, (c) 0.01, and (d) 0.02. Red bars show results from raw noisy simulations, and blue bars show results after applying error mitigation. For all error rates, the mitigated results closely recover the ideal outcome distribution, with dominant peaks at the correct bitstrings.}
	\label{fig:ghz}
\end{figure}
\rzz{Here, we provide additional results to illustrate that our method can also be applied in other types of simulations. We simulate the preparation of a five-qubit Greenberger–Horne–Zeilinger (GHZ) state using a variational quantum circuit with three layers. The target state is given by $(\ket{00000}+\ket{11111})/\sqrt{2}$. These simulations on a local simulator are presented in FIG.~\ref{fig:ghz}. Under increasing noise levels, raw noisy simulations (red bars) does display the robust GHZ signature: probability weight leaks into incorrect bitstrings, and the two dominant peaks are progressively suppressed. By contrast, simulations with error mitigation (blue bars) substantially restore the ideal distribution, recovering sharp peaks at $\ket{00000}$ and $\ket{11111}$ at increasing error rates. }

\flushbottom

\end{document}